\documentclass[preprint2,twocolappendix]{aastex6}
\usepackage{CJK}
\fullcollaborationName{The Friends of AASTeX Collaboration}

\begin{document}


\title{Resonant absorption of surface sausage and surface kink modes under photospheric conditions }


\author{Dae Jung \surname{Yu}
\altaffilmark{1,2}, Tom \surname{Van Doorsselaere}\altaffilmark{3}, and Marcel \surname{Goossens}\altaffilmark{4}}
\affil{Centre for mathematical Plasma Astrophysics, Department of Mathematics, KU Leuven\\ Celestijnenlaan
200B bus 2400, B-3001 Leuven, Belgium}

\and


\altaffiltext{1}{djyu79@gmail.com}
\altaffiltext{2}{current affiliation: School of Space Research, Kyung Hee University, Yongin 17104, Korea}
\altaffiltext{3}{tom.vandoorsselaere@kuleuven.be}
\altaffiltext{4}{marcel.goossens@kuleuven.be}

\begin{abstract}

We study the effect of resonant absorption of surface sausage and surface kink modes under photospheric conditions where the slow surface sausage modes undergo resonant damping in the slow continuum and the surface kink modes in the slow and Alfv\'{e}n continua at the transitional layers. We use recently derived analytical formulas to obtain the damping rate (time). By considering linear density and linear pressure profiles for the transitional layers, we show that resonant absorption in the slow continuum could be an efficient mechanism for the wave damping of the slow surface sausage and slow surface kink modes whilst the damping rate of the slow surface kink mode in the Alfv\'{e}n continuum is weak. It is also found that the resonant damping of the fast surface kink mode is much stronger than that of the slow surface kink mode, showing a similar efficiency as under coronal conditions. It is worth to notice that the slow body sausage and kink modes can also resonantly damp in the slow continuum for those linear profiles.

\end{abstract}





\section{Introduction}
\label{sec1}
The observed magnetohydrodynamic (MHD) waves in the solar atmosphere are considered as crucial ingredients for the coronal heating problem~\citep[e.g.][]{Ionson1978,Heyvaerts1983,Hollweg1988,Poedts1989,Poedts1990,Ofman1995,Roberts2000,Goossens2011,Okamoto2015,Antolin2017,Cally2017}.
The oscillation and rapid damping of MHD waves have made it possible to infer the physical parameters of the environment, as seismological tools. Resonant absorption has been treated as a most plausible mechanism for the rapid damping of the MHD wave oscillations and used as coronal seismology~\citep[e.g.][]{Goossens2002,Arregui2007,Goossens2008,McEwan2008,Wang2009,Wang2011,Goossens2012,Moreels2013a,
Soler2014,Moreels2015a,Moreels2015b,Wang2016,Raes2017}.

Since the energy source of the high temperature of the corona is believed to be from the convection zone below the surface of the sun, the dynamics of MHD waves in the photosphere or chromosphere is of significant interest~\citep[see e.g.][]{Jess2015,Jess2016} where sausage, kink, and torsional Alfv\'{e}n waves have mainly been investigated.

Whereas resonant absorption under coronal conditions has been extensively studied~\citep[e.g.][]{Ionson1978,Poedts1989,Ofman1995,Goossens2002,Ruderman2002,Aschwanden2003,Terradas2006a,
Terradas2006b,Ruderman2009b,Pascoe2010,Goossens2011,Soler2013,Okamoto2015,Yu2016,Scherrer2017,Karampelas2017}, its role in the lower solar atmosphere is not well understood yet~\citep{Hollweg1988,Lou1990,Rosenthal1990,Rosenthal1992,Stenuit1993,Keppens1994,Bogdan1996,Keppens1996,Ruderman2009a,Giagkiozis2016}.

In the lower atmosphere, alongside with the Alfv\'{e}n resonance, the slow (cusp) resonance can be also an important mechanism for wave energy conversion and transport.
It has been generally anticipated that the effect of resonant absorption in the slow (cusp) continuum is feeble compared to that of resonant absorption in the Alfv\'{e}n continuum~\citep[see e.g.][]{Soler2009}, which, as we recently showed in~\citet{Yu2017}, is not true for the photospheric (magnetic pore) environment. The resonant absorption mechanism may cause efficient damping of sausage modes in the photosphere, in addition to other damping effects like thermal conduction, compressive viscosity, area divergence, optically thin radiations and so on~\citep[e.g.][]{DeMoortel2003,DeMoortel2004a,Khodachenko2004,Mandal2016}.

Although we showed that the role of resonant absorption of slow surface sausage mode in the slow continuum is important for the wave damping, the model for the transitional layers was a linear cusp speed profile, which is a simple one. In this paper we put the model for transitional layer in a more general case: linear density and linear pressure (or squared magnetic field) profiles and study resonant absorption of both the surface sausage and surface kink waves under magnetic pore conditions motivated by the recent observation of the slow sausage ($m=0$) and kink ($m=1$) modes simultaneously excited in a sunspot by \citet{Jess2017}. We concentrate on the damping rate and damping time in this paper.

We organize the paper as follows. In Sec.~\ref{sec2}, we obtain the dispersion relation of surface sausage and surface kink modes under magnetic pore conditions for a plasma which is homogeneous inside and outside the pore. In Sec.~\ref{sec3}, we derive the damping rate for the slow surface waves by considering a thin transitional layer between inner and outer regions of the pore by using the connection formulae. In Sec.~\ref{sec4}, we  introduce the model configuration for the transitional layer. The results are shown in Sec.~\ref{sec5}. We conclude the paper in Sec.~\ref{sec6}.

\section{Dispersion relation}
\label{sec2}

\subsection{Dispersion relation}
\label{sec2-1}
\begin{figure}[ht]
\includegraphics[trim=10.0 100.0 25.0 30., clip, width=.55\textwidth,height=.7\textwidth]{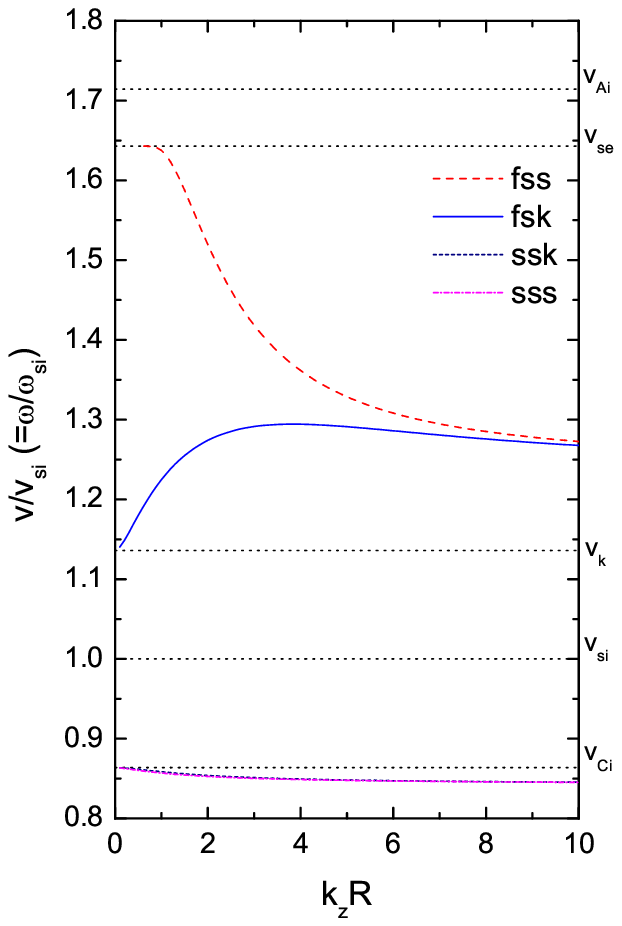}
\caption{\label{d_ph} The phase speed $v/v_{si}(=\omega_r/\omega_{si})$ as a function of $k_zR$ for a fast surface sausage mode (fss), a fast surface kink mode (fsk), a slow surface kink mode (ssk), and a slow surface sausage mode (sss) under magnetic pore condition when $v_{Ae}=0km/s$, $v_{Ai}=12km/s$, $v_{se}=11.5km/s$, $v_{si}=7km/s$, $v_{Ce}=0km/s$, $v_{Ci}\approx6.05km/s~(\approx0.86v_{si})$, $\beta_i=(2/\gamma)(v_{si}/v_{Ai})^2\approx0.41$ and $\beta_e=(2/\gamma)(v_{se}/v_{Ae})^2=\infty$. The two slow surface modes are indistinguishable in the figure. All quantities are normalized by $v_{si}$.}
\end{figure}
In our previous paper~\citep{Yu2017} we showed the dispersion relation for the fast and slow sausage modes under magnetic pore conditions by considering a uniform axisymmetric cylinder. In this paper we also consider the surface kink modes. We assume that the inside magnetic field $B_i$ and the outside magnetic field $B_e$ are parallel to the axis ($\hat{z}$) and that no steady flow is present. Then the pressures inside and outside the flux tube satisfy the pressure balance equation
\begin{eqnarray}
p_e+\frac{B_e^2}{2\mu_0}=p_i+\frac{B_i^2}{2\mu_0},\label{eq:1}
\end{eqnarray}
where $\mu_0$ is the magnetic permeability and $p$ is the plasma pressure. The subscript $i(e)$ denotes inner (outer) region of the flux tube.

We start from linearized ideal MHD equations by assuming $\exp{(i(k_zz+m\phi-\omega t))}$ dependence, where $k_z$ is the longitudinal wavenumber, $m$ the azimuthal wavenumber, and $\omega$ the angular frequency of the wave.
Here we consider no transitional layer and different physical values for the inside and outside of the flux tube boundary at $r=R$.
The density $\rho$ is assumed to be $\rho_i$ inside and $\rho_e$ outside of the boundary and the same is applied for $B$ and $p$.
The dispersion relation is then obtained by the condition of continuity at the boundary ($r=R$) \citep[e.g.][]{Edwin1983,Sakurai1991a,Goossens1992b,Yu2017}:
\begin{eqnarray}
~[P]&=&P_e-P_i=0,\label{eq:2}\\
~[\xi_r]&=&\frac{1}{\rho_e(\omega^2-\omega_{Ae}^2)}\frac{dP_e}{dr}-\frac{1}{\rho_i(\omega^2-\omega_{Ai}^2)}\frac{dP_i}{dr}=0,\label{eq:3}
\end{eqnarray}
where $P$ is the total pressure perturbation and $\xi_r$ the radial component of the Lagrangian displacement.

For the inner and outer homogeneous regions of the flux tube, the equations for $P$ and $\xi_r$ are satisfied by Bessel functions where the argument is the radial component.

For the surface wave modes, Eqs.~(\ref{eq:2}) and (\ref{eq:3}) are combined to yield~\citep[e.g.][]{Edwin1983,Yu2017}
\begin{eqnarray}
\frac{A_ek_eK_m'(k_eR)}{\rho_e(\omega^2-\omega_{Ae}^2)}-\frac{A_ik_iI_m'(k_iR)}{\rho_i(\omega^2-\omega_{Ai}^2)}=0,\label{eq:4}
\end{eqnarray}
where the prime denotes the derivative with respect to the entire argument, $I_m$ and $K_m$ are modified Bessel functions of first and second kinds respectively, $A_{i,e}$ is the matching coefficient, and $k_i$ and $k_e$ are given by
\begin{eqnarray}
k_i^2&=&-\frac{(\omega^2-\omega_{si}^2)(\omega^2-\omega_{Ai}^2)}{(v_{si}^2+v_{Ai}^2)(\omega^2-\omega_{Ci}^2)},\label{eq:5}\\
k_e^2&=&-\frac{(\omega^2-\omega_{se}^2)(\omega^2-\omega_{Ae}^2)}{(v_{se}^2+v_{Ae}^2)(\omega^2-\omega_{Ce}^2)},\label{eq:6}
\end{eqnarray}
where $\omega_{C}=k_zv_{C}$ is the cusp frequency,  $v_{C}=\sqrt{v_s^2v_A^2/(v_s^2+v_A^2)}$ the cusp speed, $\omega_s=k_zv_s$, $v_s=\sqrt{\gamma p/\rho}$ the sound speed, $v_A=B/\sqrt{\mu_0\rho}$ the Alfv\'{e}n speed, $\gamma$ the adiabatic index, and $\rho$ the density.

From the continuity of total pressure ($A_iI_m=A_eK_m$), we obtain the dispersion relation $D_m=0$ for azimuthal wavenumber $m$:
\begin{eqnarray}
D_{m}=\rho_i(\omega^2-\omega_{Ai}^2)-\rho_e(\omega^2-\omega_{Ae}^2)\bigg(\frac{k_i}{k_e}\bigg)Q_m=0,\label{eq:7}
\end{eqnarray}
where
\begin{eqnarray}
Q_m&=&\frac{I_m'(k_iR)K_m(k_eR)}{I_m(k_iR)K_m'(k_eR)}.\label{eq:8}
\end{eqnarray}
We are concerned with the sausage and kink modes ($m=0,1$) in this paper. Eq.~(\ref{eq:7}) can be rewritten as
\begin{eqnarray}
\omega^2=\frac{\rho_i\omega_{Ai}^2-\rho_e\omega_{Ae}^2\big(\frac{k_i}{k_e}\big)Q_m}{\rho_i-\rho_e\big(\frac{k_i}{k_e}\big)Q_m}~(\text{for}~ m=0,1).\label{eq:9}
\end{eqnarray}
The r.h.s. of Eq.~(\ref{eq:9}) also includes $\omega$ in $k_i$, $k_e$, and $Q_m$, therefore this equation needs to be numerically solved~\citep[see e.g.][]{Edwin1983, Yu2017}.

We use the same parameter values as in \citet{Yu2017} for the magnetic pore conditions~\citep[see also][]{Grant2015}. We plot surface wave eigenmodes for $m=0,1$ in Fig.~\ref{d_ph}: fast surface sausage mode (fss), fast surface kink mode (fsk), slow surface kink mode (ssk), and slow surface sausage mode (sss) where $v_{Ae}=0km/s$, $v_{Ai}=12km/s$, $v_{se}=11.5km/s$, $v_{si}=7km/s$, $v_{Ce}=0km/s$, and $v_{Ci}\approx6.05km/s~(\approx0.86v_{si})$. We distinguish between fast and slow mode by their phase speed: fast mode lies above the kink speed $v_k=\sqrt{(\rho_iv_{Ai}^2+\rho_ev_{Ae}^2)/(\rho_i+\rho_e)}$ and slow modes lies below $v_{Ci}$. Another characteristic is the behavior of the ratio of longitudinal to transverse component of the Lagrangian displacement $\tilde{\xi}=\xi_\parallel/\xi_\perp$ such that for the fast modes $\tilde{\xi}\lesssim1$ and for the slow modes $\tilde{\xi}>1$~\citep[see e.g.][]{Moreels2013a}. For the slow modes the longitudinal motion is dominant.

It follows from the figure that the slow surface sausage and kink modes are in the cusp frequency range ($v_{Ce}<v_{sss,ssk}\leq v_{Ci}$) while both fast and surface kink modes are in the range $v_{Ae}<v_{fsk,ssk}< v_{Ai}$. This implies that when the discontinuity is replaced by continuous variation in the transitional layers, the slow surface sausage mode (sss) and slow surface kink mode (ssk) lie in the slow (cusp) continuum and as a result damp resonantly in the resonant layer. The same phenomenon occurs for the surface kink modes (fsk, ssk) in the Alfv\'{e}n continuum. There also exist multiple body modes for sausage and kink waves in the range $v_{Ci}<v<v_{si}$, which are not shown in the figure. Since our concern is on the resonant absorption of the surface waves, we do not consider resonant absorption of the body modes here.

\subsection{Approximate dispersion relation for the slow surface kink mode at $\omega\approx\omega_{Ci}$}
\begin{figure}[ht]
\includegraphics[trim=25.0 80.0 35.0 20., clip, width=.6\textwidth]{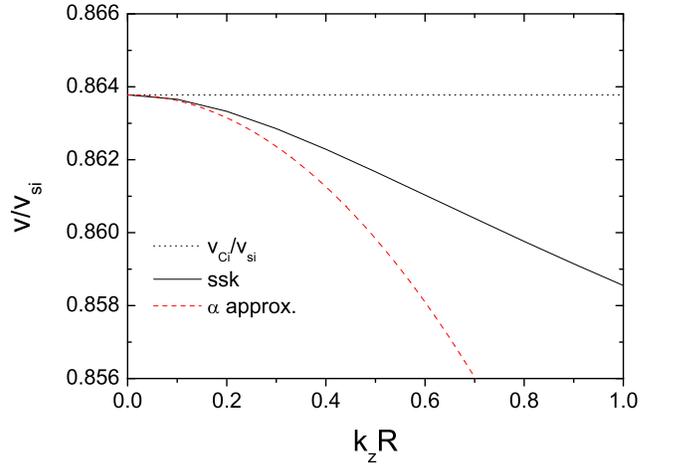}
\caption{\label{ssk} The dispersion curve as a function of $k_zR$ under magnetic pore conditions when $v_{Ae}=0km/s$, $v_{Ai}=12km/s$, $v_{se}=11.5km/s$, $v_{si}=7km/s$, $v_{Ce}=0km/s$, and $v_{Ci}\approx6.05km/s$. We compare the numerical result, Eq.~(\ref{eq:9}), with the analytical formula, Eq.~(\ref{eq:12}).}
\end{figure}
For $k_zR\ll1$ and $\omega\approx\omega_{Ci}$ we can assume $\omega^2=\omega_{Ci}^2-\alpha$, then the condition $D_{1}=0$ (Eq.~(\ref{eq:7}))
leads with the aid of Eq.~(\ref{eq:a5}) (dropping all higher order terms of $k_iR$ and $k_eR$) to
\begin{eqnarray}
\rho_i(\omega_{Ci}^2-\omega_{Ai}^2)+\rho_e(\omega_{Ci}^2-\omega_{Ae}^2)
\bigg(1+\frac{k_i^2R^2}{4}\bigg)=0.~~~~~\label{eq:10}
\end{eqnarray}
In this limit we obtain
\begin{eqnarray}
k_i^2&\approx&\frac{k_z^2}{\alpha}\frac{(\omega_{Ci}^2-\omega_{si}^2)(\omega_{Ci}^2-\omega_{Ai}^2)}
{(\omega_{si}^2+\omega_{Ai}^2)}\nonumber\\
&=&\frac{k_z^2}{\alpha}\frac{\omega_{Ci}^6}{\omega_{si}^2\omega_{Ai}^2},\label{eq:11}
\end{eqnarray}
where we have used the relations $(\omega_{Ci}^2-\omega_{si}^2)=-(\omega_{si}^2\omega_{Ci}^2)/\omega_{Ai}^2$
 and $(\omega_{Ci}^2-\omega_{Ai}^2)=-(\omega_{Ai}^2\omega_{Ci}^2)/\omega_{si}^2$.

Using Eqs.~(\ref{eq:10}) and~(\ref{eq:11}) we obtain an expression for $\alpha$ as
\begin{eqnarray}
\alpha=\frac{\chi}{4}\frac{\omega_{Ci}^6(\omega_{Ci}^2-\omega_{Ae}^2)}
{\omega_{Ci}^2\omega_{Ai}^4-\chi\omega_{si}^2\omega_{Ai}^2(\omega_{Ci}^2-\omega_{Ae}^2)}
k_z^2R^2,\label{eq:12}
\end{eqnarray}
where  $\chi=\rho_e/\rho_i=({2v_{si}^2+\gamma v_{Ai}^2})/({2v_{se}^2+\gamma v_{Ae}^2})$.

In Fig.~\ref{ssk} we compare this formula, Eq.~(\ref{eq:12}), with numerical result under magnetic pore conditions.
As shown in the figure, Eq.~(\ref{eq:12}) is accurate for $k_zR\ll 1$.

The formula for $k_i^2$ then reduces to
\begin{eqnarray}
k_i^2&=&\frac{k_z^2}{\alpha}\frac{\omega_{Ci}^6}{\omega_{si}^2\omega_{Ai}^2}\nonumber\\&=&
\frac{4}{\chi}\frac{\omega_{Ci}^2\omega_{Ai}^4-\chi\omega_{si}^2\omega_{Ai}^2(\omega_{Ci}^2-\omega_{Ae}^2)}
{\omega_{si}^2\omega_{Ai}^2(\omega_{Ci}^2-\omega_{Ae}^2)R^2}\nonumber\\&=&
\frac{4}{R^2}\bigg(\frac{\omega_{Ci}^2\omega_{Ai}^2}
{\chi\omega_{si}^2(\omega_{Ci}^2-\omega_{Ae}^2)}-1\bigg).
\label{eq:13}
\end{eqnarray}
For $k_e$ we have
\begin{eqnarray}
k_e&\approx&
k_z\sqrt{-\frac{(\omega_{Ci}^2-\omega_{se}^2)(\omega_{Ci}^2-\omega_{Ae}^2)}
{(\omega_{se}^2+\omega_{Ae}^2)(\omega_{Ci}^2-\omega_{Ce}^2)}}=k_zn_z\label{eq:14},
\end{eqnarray}
We use these formulas when we derive an analytical damping rate in the long wavelength limit.

\section{Resonant absorption due to the transitional layers}
\label{sec3}
Considering transitional layers which have a continuous variation from the inside to the outside of the flux tube, we need to solve, for example,
a second-order ordinary differential equation for $\xi_r$~\citep[e.g.][]{Sakurai1991a,Goossens1992b,Giagkiozis2016,Yu2017}:
\begin{eqnarray}
\frac{d}{dr}\bigg[\frac{D}{rC}\frac{d(r\xi_r)}{dr}\bigg]+\rho(\omega^2-\omega_A^2)\xi_r=0,\label{eq:15}
\end{eqnarray}
where
\begin{eqnarray}
D&=&\rho(\omega^2-\omega_A^2)(\omega^2-\omega_C^2)(v_s^2+v_A^2),\label{eq:16}\\
C&=&\omega^4-(v_s^2+v_A^2)(\omega^2-\omega_{C}^2)\bigg(\frac{m^2}{r}+k_z^2\bigg).\label{eq:17}
\end{eqnarray}
This differential equation has singularities at $\omega=\omega_C(r)$ and $\omega=\omega_A(r)$ where resonant absorption can occur resulting in damping of the wave amplitude. Due to the presence of the transitional layer, the value of $\omega_C(v_C)$ changes continuously from $\omega_{Ci}(v_{Ci})$ to $\omega_{Ce}(v_{Ce})$ and that of $\omega_A(v_A)$ from $\omega_{Ai}(v_{Ai})$ to $\omega_{Ae}(v_{Ae})$. These regimes are called slow (cusp) and Alfv\'{e}n continua, respectively. For the magnetic pore conditions, we obtain the relation $0=v_{Ce}=v_{Ae}<v_{Ci}<v_{si}<v_{se}<v_{Ai}$ (See~Figs.~\ref{d_ph} and ~\ref{v_p}) and no modes exist for $v>v_{se}$. The slow surface sausage mode lies in the range $0=v_{Ce}<v_{sss}<v_{Ci}$, so it can undergo resonant absorption in the slow continuum. This also applies to the slow surface kink mode since it is in the same range of the slow resonance. For the Alfv\'{e}n resonance, both the fast and slow surface kink modes lie in the range $0=v_{Ae}<v_{fsk,ssk}<v_{Ai}$, so two surface kink modes can undergo resonant absorption in the Alfv\'{e}n continuum. There is no resonant absorption of sausage modes in the Alfv\'{e}n continuum when the magnetic field is along the flux tube. The resonant absorption of the sausage mode in the Alfv\'{e}n continuum was studied by \citet{Giagkiozis2016} by considering weakly twisted magnetic flux tubes.

We are interested in the damping rate for thin transitional layers. Therefore, instead of numerically solving Eq.~(\ref{eq:15}), we use a connection formula~\citep[e.g.][]{Sakurai1991a,Goossens1992b,Soler2009,Giagkiozis2016,Yu2017}, which is demonstrated in the following sections.

{We start with a general derivation with $v_{Ae}\neq0$ and later focus on the magnetic pore conditions with $v_{Ae}=0$.}
\subsection{Connection formula}
\label{sec3-1}
As shown in Sec.~\ref{sec2} the eigenfrequency  of the slow surface sausage mode is in the slow resonance range: $\omega_r(v_{sss})<\omega_{Ci}(v_{Ci})$ and that of kink modes in the Alfv\'{e}n resonance range: $\omega_{Ae}(v_{Ae})<\omega_r(v_{fsk,ssk})<\omega_{Ai}(v_{Ai})$. Therefore these modes will undergo resonant damping in the transitional layers. When there is resonant absorption (damping) an imaginary term is included in the original dispersion relation as follows~\citep[e.g.][]{Sakurai1991a,Goossens1992b}.

Instead of the discontinuity at $r=R$, we assume a continuous variation of $\rho$ from $\rho_i$ to $\rho_e$ in a non-uniform (transitional) layer $[R- l/2, R +l/2]$ and similarly for $p$ and $B$. The thickness of the non-uniform layer is set to $l$. A fully non-uniform flux tube corresponds to $l = 2R$. By using the thin boundary approximation, we can use the analytic solutions for $P$ and $\xi_r$ in the intervals $[0, R-l/2]$ and $[R +l/2, \infty[$, avoiding numerical integration of Eq.~(\ref{eq:15}). The connection formula for $P$ is, without reference to the kind of resonance, given as
\begin{eqnarray}
~[P]&=&0,\label{eq:18}
\end{eqnarray}
which is the same as for no resonance (Eq.~(\ref{eq:2})).
Whilst the connection formula for $\xi_r$ is given as for the slow resonance
\begin{eqnarray}
~[\xi_r]&=&-i\pi\frac{k_z^2}{\rho_c|\triangle_c|}\bigg(\frac{v_{sc}^2}{v_{sc}^2+v_{Ac}^2}\bigg)^2P_c,\nonumber\\\label{eq:19}
\end{eqnarray}
where subscript $c$ denotes the position of the slow resonance ($r=r_c$) and $\triangle_c=d(\omega^2-\omega_{C}^2)/dr|_{r=r_c}$,
and for the Alfv\'{e}n resonance
\begin{eqnarray}
~[\xi_r]&=&-i\pi\frac{m^2}{\rho_A|\triangle_A|r_A^2}P_A~,\label{eq:20}
\end{eqnarray}
where subscript $A$ denotes the position of the Alfv\'{e}n resonance ($r=r_A$) and $\triangle_A=d(\omega^2-\omega_{A}^2)/dr|_{r=r_A}$.

From Eqs.~(\ref{eq:3}),~(\ref{eq:19}), and (\ref{eq:20}) we obtain
\begin{eqnarray}
\frac{P_e'}{\rho_e(\omega^2-\omega_{Ae}^2)}&-&\frac{P_i'}{\rho_i(\omega^2-\omega_{Ai}^2)}=\nonumber\\
&-&i\pi\frac{k_z^2}{\rho_c|\triangle_c|}\bigg(\frac{v_{sc}^2}{v_{sc}^2+v_{Ac}^2}\bigg)^2P_c,\label{eq:21}
\end{eqnarray}
and
\begin{eqnarray}
\frac{P_e'}{\rho_e(\omega^2-\omega_{Ae}^2)}&-&\frac{P_i'}{\rho_i(\omega^2-\omega_{Ai}^2)}=-i\pi\frac{m^2}{\rho_A|\triangle_A|r_A^2}P_A.
\nonumber\\\label{eq:22}
\end{eqnarray}

\subsection{Analytical solution for the damping rate of the slow surface modes of sausage and kink waves in the slow continuum}
\label{sec3-2}
In our previous paper~\citep{Yu2017} we have developed an analytical formula for the damping rate of the slow mode in the slow continuum. Here we introduce again the procedure for obtaining the damping rate.
For the surface mode, Eq.~(\ref{eq:21}) can be reduced to
\begin{eqnarray}
\frac{A_ek_eK_m'(k_eR)}{\rho_e(\omega^2-\omega_{Ae}^2)}&-&\frac{A_ik_iI_m'(k_iR)}{\rho_i(\omega^2-\omega_{Ai}^2)}\label{eq:23}\\&+&
\frac{i\pi k_z^2}{\rho_c|\triangle_c|}\bigg(\frac{v_{sc}^2}{v_{sc}^2+v_{Ac}^2}\bigg)^2A_eK_m(k_eR)=0,\nonumber
\end{eqnarray}
where we have used the continuity of $P$ ($P_i=P_e=P_c$) and $A_{i,e,c}$ is the matching coefficient.

As before for the discontinuous case, we can eliminate the coefficients $A_i$, $A_e$ to arrive at the dispersion relation. The dispersion function $D_m$ has a real and an imaginary part: $D_m=D_{mr}+iD_{mi}$. Eliminating the matching coefficients by using the continuity of the total pressure, we have the dispersion relation for $D_m=0$
\begin{eqnarray}
&&\rho_i(\omega^2-\omega_{Ai}^2)-\rho_e(\omega^2-\omega_{Ae}^2)\frac{k_i}{k_e}Q_m\nonumber\\&&+
\frac{i\pi k_z^2}{\rho_c|\triangle_c|}\bigg(\frac{v_{sc}^2}{v_{sc}^2+v_{Ac}^2}\bigg)^2
\rho_i\rho_e(\omega^2-\omega_{Ai}^2)(\omega^2-\omega_{Ae}^2)\frac{G_m}{k_e}=0,\nonumber\\\label{eq:24}
\end{eqnarray}
where
\begin{eqnarray}
G_m=\frac{K_m(k_eR)}{K_m'(k_eR)}.\label{eq:25}
\end{eqnarray}
We define $D_{mi}$ as
\begin{eqnarray}
D_{mi}=\frac{\pi \rho_i\rho_e k_z^2}{k_e\rho_c|\triangle_c|}\bigg(\frac{v_{sc}^2}{v_{sc}^2+v_{Ac}^2}\bigg)^2
(\omega^2-\omega_{Ai}^2)(\omega^2-\omega_{Ae}^2)G_m,\nonumber\\\label{eq:26}
\end{eqnarray}
and $D_{mr}$ as
\begin{eqnarray}
D_{mr}=\rho_i(\omega^2-\omega_{Ai}^2)-\rho_e(\omega^2-\omega_{Ae}^2)\bigg(\frac{k_i}{k_e}\bigg)Q_m,\label{eq:27}
\end{eqnarray}
which is the same as Eq.~(\ref{eq:7}).

Due to resonant damping the wave frequency has a real and an imaginary part: $\omega=\omega_r+i\gamma_m$. The imaginary part $\gamma_m$ can be obtained by $\gamma_m=-D_{mi}/(\partial D_{mr}/\partial \omega)|_{\omega=\omega_r}$ \citep[e.g.][]{Krall1973,Goossens1992b} by assuming $|\gamma_m|\ll \omega_r$.

The analytical formula for $\gamma_m$  (see Appendix~\ref{append-2}) in the slow (cusp) continuum is given as
\begin{eqnarray}
\gamma_{cm}&=&-\frac{\frac{\pi \rho_e k_z^2}{k_e\rho_c|\triangle_c|}\bigg(\frac{v_{sc}^2}{v_{sc}^2+v_{Ac}^2}\bigg)^2 
(\omega_r^2-\omega_{Ai}^2)(\omega_r^2-\omega_{Ae}^2)G_m}
{2\omega_r\big[1-\chi\big(\frac{k_i}{k_e}\big)Q_m\big]
-\omega_r\chi T_m},~~~~~~~\label{eq:28}
\end{eqnarray}
where
\begin{eqnarray}
T_m&=&\omega_r^2(\omega_r^2-\omega_{Ae}^2)\bigg(\frac{k_i}{k_e}\bigg)\bigg\{
\frac{(\omega_r^2-2\omega_{Ci}^2)[Q_m+k_iR P_m]}{(\omega_r^2-\omega_{si}^2)(\omega_r^2-\omega_{Ai}^2)(\omega_r^2-\omega_{Ci}^2)}\nonumber\\
&&-\frac{(\omega_r^2-2\omega_{Ce}^2)[Q_m-k_eR S_m]}{(\omega_r^2-\omega_{se}^2)
(\omega_r^2-\omega_{Ae}^2)(\omega_r^2-\omega_{Ce}^2)}\bigg\},\label{eq:29}\\
P_m&=&\bigg(\frac{I_m''}{I_m}-\frac{I_m'^2}{I_0^2}\bigg)\frac{K_m}{K_m'},\label{eq:30}\\
S_m&=&\bigg(1-\frac{K_m''K_m}{K_m'^2}\bigg)\frac{I_m'}{I_m}\label{eq:31}.
\end{eqnarray}

For the sausage ($m=0$) and kink ($m=1$) modes in the slow continuum we obtain
\begin{eqnarray}
\gamma_{c0}&=&-\frac{\pi \rho_e k_z^2}{k_e\rho_c|\triangle_c|}\bigg(\frac{v_{sc}^2}{v_{sc}^2+v_{Ac}^2}\bigg)^2
\frac{(\omega_r^2-\omega_{Ai}^2)(\omega_r^2-\omega_{Ae}^2)G_0}
{2\omega_r\big[1-\chi\big(\frac{k_i}{k_e}\big)Q_0\big]
-\omega_r\chi T_0},\label{eq:32}\nonumber\\
\end{eqnarray}
and
\begin{eqnarray}
\gamma_{c1}&=&-\frac{\pi \rho_e k_z^2}{k_e\rho_c|\triangle_c|}\bigg(\frac{v_{sc}^2}{v_{sc}^2+v_{Ac}^2}\bigg)^2
\frac{(\omega_r^2-\omega_{Ai}^2)(\omega_r^2-\omega_{Ae}^2)G_1}
{2\omega_r\big[1-\chi\big(\frac{k_i}{k_e}\big)Q_1\big]
-\omega_r\chi T_1}.\label{eq:33}\nonumber\\
\end{eqnarray}
{A nonzero value of $v_{Ae}$ has opposite effects on $\gamma$, since it decreases the value of $\omega^2_{Ci}-\omega^2_{Ae}$ in the numerator and decreases the value of $|\Delta_c|$ in the denominator, which may increase or decrease the damping rate depending on the variation of the two factors. This equally applies to the damping in the Alfv\'{e}n resonance.}
\subsection{Long wavelength limit $(m=0,1)$}
\label{sec3-3}
 We derived previously an analytical expression for the damping rate of slow surface sausage mode in the slow continuum in the long wavelength limit~\citep{Yu2017}. Here we describe the result briefly.
In the limit $k_iR(k_eR)\ll1$, we obtain
\begin{eqnarray}
\gamma_{c0}&=&-\frac{\pi \rho_i\rho_e k_z^2R}{\rho_c|\triangle_c|}\bigg(\frac{v_{sc}^2}{v_{sc}^2+v_{Ac}^2}\bigg)^2\nonumber\\
&&\times\frac{(\omega_r^2-\omega_{Ai}^2)(\omega_r^2-\omega_{Ae}^2)\ln(k_eR)}                
{2\omega_r\bigg\{\rho_i-\frac{\rho_ek_i^2R^2}{2}\ln(k_eR)\bigg\}
-\rho_e\omega_rT_0},\nonumber\\\label{eq:34}
\end{eqnarray}
where
\begin{eqnarray}
T_0&=&\omega_r^2(\omega_r^2-\omega_{Ae}^2)\bigg\{
\frac{3}{16}\frac{(\omega_r^2-2\omega_{ci}^2)k_i^4R^4\ln(k_eR)}
{(\omega_r^2-\omega_{si}^2)(\omega_r^2-\omega_{Ai}^2)(\omega_r^2-\omega_{Ci}^2)}\nonumber\\
&&+\frac{(\omega_r^2-2\omega_{Ce}^2)k_i^2R^2}{2(\omega_r^2-\omega_{se}^2)
(\omega_r^2-\omega_{Ae}^2)(\omega_r^2-\omega_{Ce}^2)}\bigg\}.\label{eq:35}
\end{eqnarray}
When $\omega_r\approx\omega_{Ci}~(k_zR\ll1)$  Eq.~(\ref{eq:34}) becomes
\begin{eqnarray}
\gamma_{c0}&=&\frac{2\pi\chi^3}{3|\triangle_c|R}
\frac{\omega_{Ci}^5\omega_{si}^2(\omega_{Ci}^2-\omega_{Ae}^2)^3}                
{\omega_{Ai}^{10}}(k_zR)^4\ln^3(k_zR),\nonumber\\\label{eq:36}
\end{eqnarray}
{where we have used $\ln(k_eR)\approx\ln(k_zR)$.}

Under photospheric (magnetic pore) conditions where $\omega_{Ae}(\omega_{Ce})\simeq0$, Eq.~(\ref{eq:36}) can be reduced to
\begin{eqnarray}
\gamma_{c0}&=&\frac{2\pi\chi^3}{3|\triangle_c|R}
\frac{\omega_{Ci}^{11}\omega_{si}^2}
{\omega_{Ai}^{10}}(k_zR)^4\ln^3(k_zR).
\label{eq:37}
\end{eqnarray}
Likewise, using the approximations for $G_1$, $Q_1$, $T_1$ used in Sec.~\ref{sec3-5}, we find for Eq.~(\ref{eq:33}) under magnetic pore conditions
\begin{eqnarray}
\gamma_{c1}&\approx&-\frac{\pi }{8|\triangle_c|R}
\frac{\omega_{Ci}^7k_z^4R^4}{\omega_{si}^4}\frac{{\omega_{Ci}}^4}{\omega_{Ai}^4}
\bigg(\frac{\omega_{Ai}^2}{\chi\omega_{si}^2}-1\bigg)^{-2}\nonumber\\
&=&-\frac{\pi }{8|\triangle_c|R}
\frac{\chi^2\omega_{Ci}^{11}k_z^4R^4}{\omega_{Ai}^4(\omega_{Ai}^2-\chi\omega_{si}^2)^2}.
\label{eq:38}
\end{eqnarray}
For comparison of Eqs.~(\ref{eq:37}) and ~(\ref{eq:38}), see Eq.~(\ref{eq:50}).

\subsection{Analytical solution for the damping rate of the surface kink mode in the Alfv\'{e}n continuum}
\label{sec3-4}
Considering the Alfv\'{e}n resonance, we have a dispersion relation from Eq.~(\ref{eq:22})
\begin{eqnarray}
&&\rho_i(\omega^2-\omega_{Ai}^2)-\rho_e(\omega^2-\omega_{Ae}^2)\frac{k_i}{k_e}Q_m\nonumber\\&&+
\frac{i\pi m^2}{\rho_A|\triangle_A|r_A^2}
\rho_i\rho_e(\omega^2-\omega_{Ai}^2)(\omega^2-\omega_{Ae}^2)\frac{G_m}{k_e}=0,\nonumber\\\label{eq:39}
\end{eqnarray}
and the analytical formula for $\gamma_m$ is given as
\begin{eqnarray}
\gamma_{Am}&=&-\frac{\frac{\pi m^2\rho_e }{k_e\rho_A|\triangle_A|r_A^2} 
(\omega_r^2-\omega_{Ai}^2)(\omega_r^2-\omega_{Ae}^2)G_m}
{2\omega_r\big[1-\chi\big(\frac{k_i}{k_e}\big)Q_m\big]
-\omega_r\chi T_m},\label{eq:40}
\end{eqnarray}
where only the numerator is slightly changed when compared with Eq.~\ref{eq:28}. From Eq.~(\ref{eq:39}), it is inferred that no resonant absorption in the Alfv\'{e}n continuum occurs for the sausage waves since the imaginary part becomes zero when $m=0$.

 For the surface kink mode ($m=1$) in the Alfv\'{e}n continuum we obtain
\begin{eqnarray}
\gamma_{A1}&=&-\frac{\pi \rho_e }{k_e\rho_A|\triangle_A|r_A^2}
\frac{(\omega_r^2-\omega_{Ai}^2)(\omega_r^2-\omega_{Ae}^2)G_1}
{2\omega_r\big[1-\chi\big(\frac{k_i}{k_e}\big)Q_1\big]
-\omega_r\chi T_1}.\nonumber\\\label{eq:41}
\end{eqnarray}

\subsection{Long wavelength limit $(m=1)$}
\label{sec3-5}
In the limit $k_iR(k_eR)\ll1$, $\gamma_{A1}$ reduces to (see Appendix \ref{append-2}), by using the asymptotic expansion of $Q_1$, $G_1$, $P_1$ and $S_1$ (Eqs. (\ref{eq:a5}-\ref{eq:a8})),
\begin{eqnarray}
\gamma_{A1}&=&\frac{\pi \rho_e R}{\rho_A|\triangle_A|r_A^2}\nonumber\\
&&\times\frac{(\omega_r^2-\omega_{Ai}^2)(\omega_r^2-\omega_{Ae}^2)}                
{2\omega_r\bigg[1+\chi\bigg(1+\frac{k_i^2R^2}{4}\bigg)\bigg]
-\omega_r\chi T_1},\nonumber\\\label{eq:42}
\end{eqnarray}
where
\begin{eqnarray}
T_1&=&-\omega_r^2(\omega_r^2-\omega_{Ae}^2)\bigg\{
\frac{(\omega_r^2-2\omega_{Ci}^2)(k_iR)^2}{2(\omega_r^2-\omega_{si}^2)(\omega_r^2-\omega_{Ai}^2)(\omega_r^2-\omega_{Ci}^2)}\nonumber\\&&
+\frac{(\omega_r^2-2\omega_{Ce}^2)[-(k_iR)^2/4+(1+3\ln(k_eR))(k_eR)^2]}{(\omega_r^2-\omega_{se}^2)
(\omega_r^2-\omega_{Ae}^2)(\omega_r^2-\omega_{Ce}^2)}\bigg\}.\nonumber\\\label{eq:43}
\end{eqnarray}
For $k_zR\ll1$  and $\omega_r\approx\omega_{Ci}$ (using Eqs.~(\ref{eq:12})-(\ref{eq:14})), Eq.~(\ref{eq:42}) is reduced to
\begin{eqnarray}
\gamma_{A1}&=&\frac{\pi\chi R}{|\triangle_A|r_i^2}
\frac{(\omega_{Ci}^2-\omega_{Ai}^2)(\omega_{Ci}^2-\omega_{Ae}^2)}                
{2\omega_{Ci}\bigg[1+\frac{\omega_{Ci}^2\omega_{Ai}^2}
{\omega_{si}^2(\omega_{Ci}^2-\omega_{Ae}^2)}\bigg]
-\chi\omega_{Ci}T_1},\nonumber\\\label{eq:44}
\end{eqnarray}
where
\begin{eqnarray}
T_1&=&-\frac{8\omega_{si}^2\omega_{Ai}^2(\omega_{Ci}^2-\omega_{Ae}^2)}{\omega_{Ci}^6k_z^2R^2}
\bigg(\frac{\omega_{Ci}^2\omega_{Ai}^2}{\chi\omega_{si}^2(\omega_{Ci}^2-\omega_{Ae}^2)}-1\bigg)^2,~~~~~~\label{eq:45}
\end{eqnarray}
where we have left the most dominant term (see Appendix \ref{append-2}). Due to the factor $1/(k_zR)^2$ in the denominator of Eq.~(\ref{eq:45}), we may further reduce Eq.~(\ref{eq:44}) into
\begin{eqnarray}
\gamma_{A1}&\approx&-\frac{\pi }{8|\triangle_A|R}
\frac{\omega_{Ci}^7k_z^2R^2}{\omega_{si}^4}
\bigg(\frac{\omega_{Ci}^2\omega_{Ai}^2}{\chi\omega_{si}^2(\omega_{Ci}^2-\omega_{Ae}^2)}-1\bigg)^{-2}
,\nonumber\\\label{eq:46}
\end{eqnarray}
where we have used $r_i\approx R$.
For the photospheric (magnetic pore) conditions ($\omega_{Ae},\omega_{Ce}\simeq0$), we obtain
\begin{eqnarray}
\gamma_{A1}&=&-\frac{\pi }{8|\triangle_A|R}
\frac{\omega_{Ci}^7k_z^2R^2}{\omega_{si}^4}
\bigg(\frac{\omega_{Ai}^2}{\chi\omega_{si}^2}-1\bigg)^{-2}\nonumber\\
&=&-\frac{\pi }{8|\triangle_A|R}
\frac{\chi^2\omega_{Ci}^{7}k_z^2R^2}{(\omega_{Ai}^2-\chi\omega_{si}^2)^2}.
\label{eq:47}
\end{eqnarray}

For two slow surface modes in the long wavelength limit, comparison of the above three resonant absorption effects leads to the conclusion that the wave damping due to the Alfv\'{e}n resonance is stronger than that due to the slow resonance:
\begin{eqnarray}
\frac{\gamma_{A1}}{\gamma_{c0}}&=&-\frac{3}{16\chi}\frac{|\triangle_c|}{|\triangle_A|}
\frac{\omega_{Ai}^{10}}{\omega_{si}^2\omega_{Ci}^4(\omega_{Ai}^2-\chi\omega_{si}^2)^2}
\frac{1}{k_z^2R^2\ln^3(k_zR)},\nonumber\\\label{eq:48}\\
\frac{\gamma_{A1}}{\gamma_{c1}}&=&\frac{|\triangle_c|}{|\triangle_A|}\frac{\omega_{Ai}^4}{\omega_{Ci}^4k_z^2R^2},\label{eq:49}\\
\frac{\gamma_{c0}}{\gamma_{c1}}&=&-\frac{16\chi}{3}\frac{\omega_{si}^2(\omega_{Ai}^2-\chi\omega_{si}^2)^2\ln^3(k_zR)}{\omega_{Ai}^6}.\label{eq:50}
\end{eqnarray}
These formulas provide a relative strength among three different resonant absorptions for the surface sausage and kink modes in the long wavelength limit such that at $k_zR\approx0$ the damping due to the Alfv\'{e}n resonance is much stronger than due to the slow resonance and, for the slow resonance, the resonant absorption for the slow surface sausage mode is more stronger than for the slow surface kink mode. These features are proven in Fig.~\ref{dp_sk1}. In the figure, it is also shown that there is a crossover between two curves for slow resonance and the curve for Alfv\'{e}n resonance, after which the resonant damping of the slow resonance dominates over that of the Alfv\'{e}n resonance.

{Caution is needed for using these formulae in the long wavelength limit, given their limited validity range, as we showed in Fig. 5 in \citet{Yu2017}.}

\begin{figure}[ht]
\includegraphics[trim=20.0 80.0 45.0 30., clip, width=.6\textwidth]{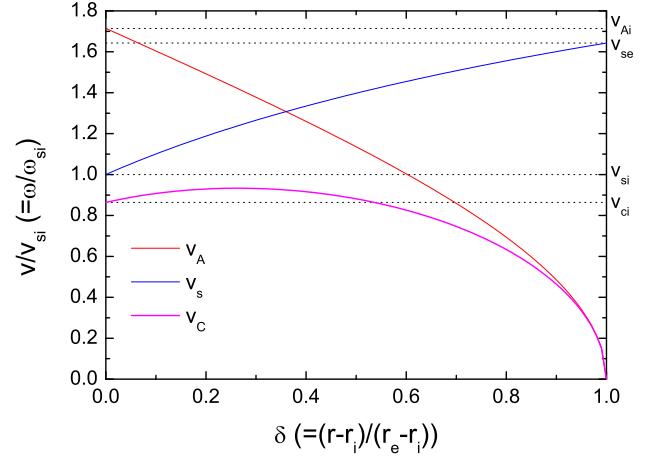}
\caption{\label{v_p} The profiles for $v_{s}$, $v_{A}$ and $v_{C}$ as a function of $\delta$ in the non-uniform (transitional) layer under magnetic pore conditions when $v_{Ae}=0km/s$, $v_{Ai}=12km/s$, $v_{se}=11.5km/s$, $v_{si}=7km/s$, $v_{Ce}=0km/s$, and $v_{Ci}\approx6.05km/s$. When $v_{Ci}<v<v_{Cm}$ the slow body sausage modes can resonantly damp in the slow continuum where $v_{Cm}$ is the maximum value of $v_C$. When $v<v_{Ci}$ the slow surface sausage mode can resonantly damp in the slow continuum. The slow surface kink mode may resonantly damp both in the slow continuum and in the Alfv\'{e}n continuum whilst the fast surface kink mode undergoes resonant absorption in the Alfv\'{e}n continuum. There is no resonant absorption for the sausage modes in the Alfv\'{e}n continuum when the external magnetic field is along the tube axis (no azimuthal component) as considered in this paper.}
\end{figure}
\section{Linear profiles for the density and pressure}
\label{sec4}

In this paper we consider a linear profile for the density and pressure (or equivalently squared magnetic field) in the non-uniform layer.
For the linear density profile we define $\rho=\rho_i+(\rho_e-\rho_i)(r-r_i)/(r_e-r_i)$. The position of resonance where resonant absorption occurs depends on the wave frequency in the slow or Alfv\'{e}n continuum ($v=v_{C,A}$): $r=r(v)$. We introduce a new variable $\delta$ such that  $r=r_i+\delta(r_e-r_i)$ in the transitional layers where $0\leq\delta\leq1$. That is $\delta=(r-r_i)/(r_e-r_i)$. This makes it more convenient to derive the formula for the position of resonance in terms of resonance (cusp or Alfv\'{e}n) frequency. Then we can represent the density $\rho$ as a function of $\delta$ such that $\rho=\rho_i+\delta(\rho_e-\rho_i)$. Assuming also a linear variation of pressure $p$ we may set $p=p_i+\delta(p_e-p_i)$ as like  $\rho$, then it is straightforward to show that $B^2$ also has a similar relation $B^2=B_i^2+\delta(B_e^2-B_i^2)$. In this way, the variables $v_{s}$, $v_{A}$, and $v_{C}$ can be represented as (see Appendix \ref{append-3})
\begin{eqnarray}
v_{s}&=&v_{si}\frac{\sqrt{1+\delta(\chi v_{sei}^2-1)}}{\sqrt{1+\delta(\chi-1)}},\label{eq:51}\\
v_{A}&=&v_{Ai}\frac{\sqrt{1+\delta(\chi v_{Aei}^2-1)}}{\sqrt{1+\delta(\chi-1)}},\label{eq:52}\\
v_{C}^2&=&\frac{v_{si}^2v_{Ai}^2}{v_{si}^2[1+\delta(\chi v_{sei}^2-1)]+v_{Ai}^2[1+\delta(\chi v_{Aei}^2-1)]}
\nonumber\\&&\times\frac{[1+\delta(\chi v_{sei}^2-1)][1+\delta(\chi v_{Aei}^2-1)]}{1+\delta(\chi-1)},\label{eq:53}
\end{eqnarray}
where $v_{sei}^2=v_{se}^2/v_{si}^2$ and $v_{Aei}^2=v_{Ae}^2/v_{Ai}^2$.

In Fig.~\ref{v_p} we plot $v_s$, $v_A$, and $v_C$ under the magnetic pore condition where $v_{Ae}=0km/s$, $v_{Ai}=12km/s$, $v_{se}=11.5km/s$, and $v_{si}=7km/s$. The parameters inside the magnetic pore are taken from \cite{Grant2015} and those outside the magnetic pore are typical values of the photosphere. Note that the cusp speed $v_C$ covers some range of slow body and slow surface modes, so resonant absorption can occur for both slow surface and slow body sausage modes in the slow continuum and for slow body kink modes in the Alfv\'{e}n continuum.

Since the value of $\delta$ is not obtainable from observations with current resolution of the instruments, we need to find the expression for $\delta$ in terms of $v_C$ or $v_A$~\citep[see e.g.][]{Soler2009}. From Eq.~(\ref{eq:53}) we derive the quadratic formula for $\delta(=\delta_c)$ with respect to $v_{C}$
\begin{eqnarray}
A\delta^2+B\delta+C=0,\label{eq:54}
\end{eqnarray}
where
\begin{eqnarray}
A&=&1+\frac{v_C^2}{v_{Ci}^2}(\chi-1)+\chi\bigg[\frac{v_C^2}{v_{Ci}^2}-(v_{sei}^2+v_{Aei}^2)\bigg]
\nonumber\\&&-\chi^2\bigg(\frac{v_C^2}{v_{Ci}^2}-v_{sei}^2v_{Aei}^2\bigg),\label{eq:55}\\
B&=&2\bigg(\frac{v_C^2}{v_{Ci}^2}-1\bigg)-\chi\bigg[\frac{v_C^2}{v_{Ci}^2}
\bigg(1+\frac{v_{se}^2+v_{Ae}^2}{v_{si}^2+v_{Ai}^2}\bigg)\nonumber\\&&-(v_{sei}^2+v_{Aei}^2)\bigg],\label{eq:56}\\
C&=&1-\frac{v_C^2}{v_{Ci}^2},\label{eq:57}
\end{eqnarray}
which yields two solutions (see the curve $v_C$ in Fig.~\ref{v_p}):
\begin{eqnarray}
\delta_{c1}&=&-\frac{B}{2A}+\frac{\sqrt{B^2-4AC}}{2A}~~(0<\delta_{c1}\leq\delta_m),\label{eq:58}\\
\delta_{c2}&=&-\frac{B}{2A}-\frac{\sqrt{B^2-4AC}}{2A}~~(\delta_m<\delta_{c2}\leq1),\label{eq:59}
\end{eqnarray}
where $\delta_m$ is the value of $\delta$ when $v$ has a maximum value $v_{Cm}$ (here $\delta_m\approx0.26$, $v_{Cm}\approx0.93v_{si}$ for the magnetic pore condition).
When $0<\delta(=\delta_{c1})<\delta_m$ $v_{C}$ is from $v_{Ci}$ to $v_{Cm}$. When $\delta_m<\delta(=\delta_{c2})<1$, $v_C$ is from $v_{Cm}$ to $v_{Ce}$.

As a result, $\triangle_c$ is given as (see Appendix \ref{append-3})
    \begin{eqnarray}
\triangle_c&=&-\bigg(\frac{\omega_{C}^2}{l}\bigg)
\bigg\{\frac{(\chi v_{sei}^2-1)}{1+\delta_c(\chi v_{sei}^2-1)}-\frac{(\chi-1)}{1+\delta_c(\chi-1)}\nonumber\\&&+\frac{(\chi v_{Aei}^2-1)}{1+\delta_c(\chi v_{Aei}^2-1)}\label{eq:60}\\
&&-\frac{v_{si}^2(\chi v_{sei}^2-1)+v_{Ai}^2(\chi v_{Aei}^2-1)}{v_{si}^2[1+\delta_c(\chi v_{sei}^2-1)]+v_{Ai}^2[1+\delta_c(\chi v_{Aei}^2-1)]}\bigg\},\nonumber
\end{eqnarray}
where $\omega_C=\omega_C(\delta=\delta_c)$ and $\delta_c=\delta_{c1},\delta_{c2}$

For the slow surface sausage mode to resonantly damp, $v(=\omega_r/k_z)$ should be below $v_{Ci}$, which means that only $\delta_{c2}$ satisfies this condition. For the slow body sausage modes to undergo resonant damping, both solutions are needed because for $v_{Ci}<v<v_{Cm}$ resonant absorption occurs at two resonance positions $\delta_{c1}$ and $\delta_{c2}$.

From Eq. (\ref{eq:52}) we derive a formula for $\delta(=\delta_a)$ with respect to $v_{A}$
\begin{eqnarray}
\delta_a=\frac{1-(v_A/v_{Ai})^2}{1-(v_A/v_{Ai})^2+\chi[(v_A/v_{Ai})^2-v_{Aei}]}.\label{eq:61}
\end{eqnarray}

Then $\triangle_A$ becomes (see Appendix \ref{append-3})
\begin{eqnarray}
\triangle_A&=&-\bigg(\frac{\omega_{A}^2}{l}\bigg)
\bigg\{\frac{\chi v_{Aei}^2-1}{1+\delta_a(\chi v_{Aei}^2-1)}-\frac{\chi-1}{1+\delta_a(\chi-1)}\bigg\},\nonumber\\\label{eq:62}
\end{eqnarray}
where $\omega_A=\omega_A(\delta=\delta_a)$.
The resonant position $r_A$ can be written in terms of $\delta_a$ as $r_A=R+l(\delta_a-0.5)$, which we use in the calculation of Eq.~(\ref{eq:41}) and Eq.~(\ref{eq:44}).


%

\begin{figure}[ht]
\includegraphics[trim=5.0 85.0 5.0 10., clip, width=.55\textwidth]{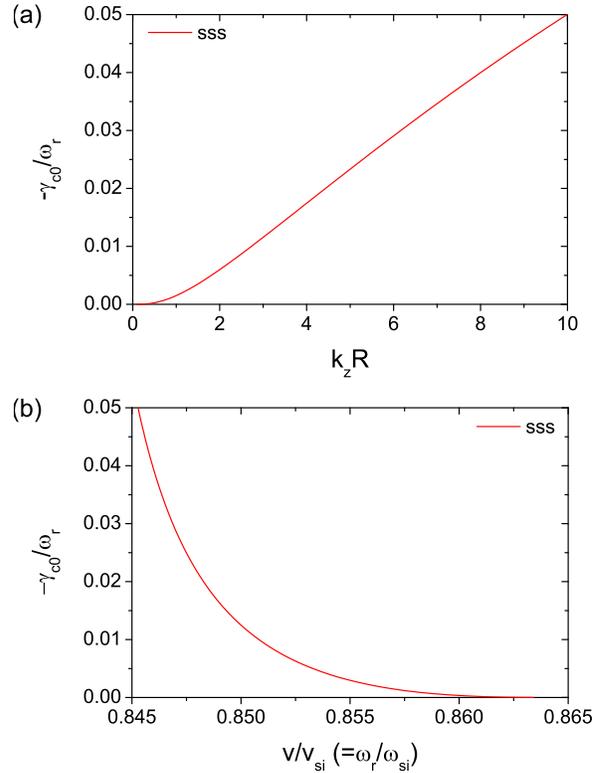}
\caption{\label{dp_sss1} The analytical formula for the damping rate $-\gamma_{c0}/\omega_r$, Eq.~(\ref{eq:32}), versus (a) $k_zR$ and (b) $v/v_{si}(=\omega/\omega_{si})$ for the slow sausage surface mode (sss) where $l/R=0.1$, $v_{Ae}=0km/s$, $v_{Ai}=12km/s$, $v_{se}=11.5km/s$, $v_{si}=7km/s$, $\beta_i=(2/\gamma)(v_{si}/v_{Ai})^2=0.4083$ and $\beta_e=(2/\gamma)(v_{se}/v_{Ae})^2=\infty$. The linear profiles for the density and pressure considered in Sec.~\ref{sec4} are used. }
\end{figure}
\section{Results}
\label{sec5}

\begin{figure}[ht]
\includegraphics[trim=5.0 85.0 20.0 10., clip, width=.5\textwidth]{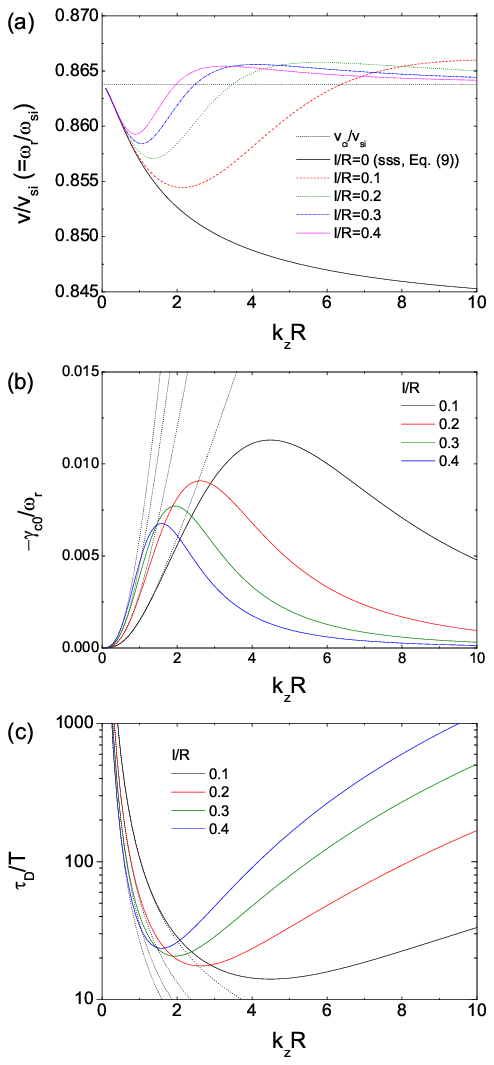}
\caption{\label{dp_sssn1} (a) The phase speed of the slow surface sausage (sss) mode $v/v_{si}$ versus $k_zR$. We compare the solution of Eq.~(\ref{eq:9}) (solid line) without an inhomogeneous (transitional) layer with the solutions (dashed, dotted, dashed-dotted, and short-dashed lines) of Eq.~(\ref{eq:24}) with the transitional layer introduced in Sec.~\ref{sec4} when $l/R=0.1,0.2,0.3,0.4$. {The other parameters are the same as in previous figures.} Each solution curve above the dotted line ($v_{Ci}$) corresponds to the one of body modes. (b) The damping rate $-\gamma_{c0}/\omega_r$ versus $k_zR$ corresponding to curves in (a). The analytical approximations (dotted lines), Eq.~(\ref{eq:32}), are compared to the numerical solutions (solid lines), Eq.~(\ref{eq:24}), for $m=0$. As $l/R$ increases, the curve of the numerical solution shifts to the left while decreasing. The analytical and numerical solutions for each $l/R$ converge when the value of $k_zR$ approach zero. (c) The ratio of the damping time to the period $\tau_D/T$ (logarithmic scale) versus $k_zR$. The position of the dip shifts to the right as $l/R$ decreases and its value tends to approximately approach 10.5. }
\end{figure}
\begin{figure}[ht]
\includegraphics[trim=5.0 90.0 20.0 10., clip, width=.5\textwidth]{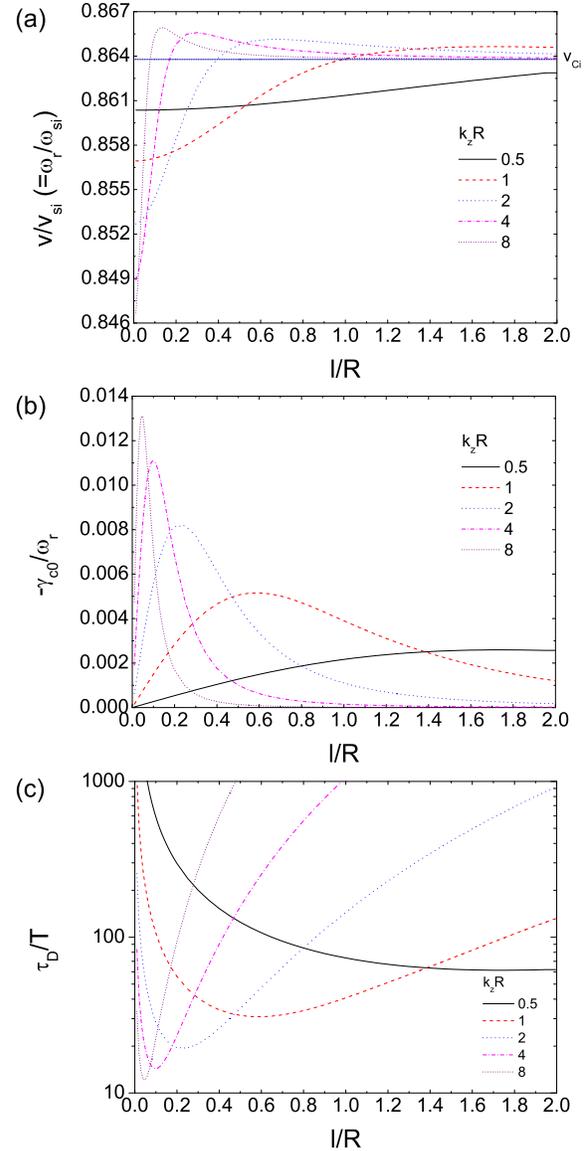}
\caption{\label{dp_sssn2} (a) $v/v_{si}$ versus $l/R$ for the slow surface sausage (sss) mode when $k_zR=0.5,1,2,4,8$. The other parameters are the same as in previous figures. Each solution curve above the dotted line ($v_{Ci}$) corresponds to the one of body modes. For larger $k_zR$, as $l/R$ increases the wave frequency approaches $\omega_{Ci}(v_{Ci})$. (b) The damping rate $-\gamma_{c0}/\omega_r$ versus $l/R$. Each curve has a local peak whose position shifts to a smaller $l/R$ as $k_zR$ increases, resulting in higher damping rate. (c) $\tau_D/T$ (logarithmic scale) versus $l/R$. For sufficiently large value of $k_zR$, it has a local dip which moves to smaller $l/R$ as $k_zR$ increases. }
\end{figure}

We have considered the linear density and linear pressure (squared magnetic field) profiles for the transitional layer given in Sec.~\ref{sec4}. We first deal with resonant absorption in the slow continuum. In Fig.~\ref{dp_sss1}, by using an analytical formula, Eq.~(\ref{eq:32}), we plot the damping rate $-\gamma_{c0}/\omega_r$ for the slow surface sausage (sss) mode as a function of (a) $k_zR$ and (b) $v/v_{si}$ when $l/R=0.1$. The parameters for each phase speed are described in the caption. The damping rate $-\gamma_{c0}/\omega_r$ increases as $k_zR$ increases and as $v/v_{si}$ decreases. If we take $k_zR=5$, $-\gamma_{c0}/\omega_r\approx0.023$, then the ratio of damping time to the period $\tau_D/T$ is $(1/|\gamma_{c0}|)/(2\pi/\omega_r)=1/(2\pi |\gamma_{c0}|/\omega_r)\approx 6.825$, which is a bit larger than the typical value for the resonant damping of the kink mode (2-4). This result could mean, contrary to previous interpretation, that the slow continuum may play a key role in the decay of the slow sausage mode and heating the lower chromosphere in certain situations. Although Eq.~(\ref{eq:32}) is valid for a small damping ($|\gamma_{m}|\ll\omega_r$), it is necessary to check its validity range by comparing with the numerical solution of Eq.~(\ref{eq:24}).

We compare the above analytical result with the numerical result. To obtain the analytical solution (Eq.~(\ref{eq:32})) we previously put $\omega_r(v)$ equal to the eigenfrequency of the undamped situation (i.e. $l/R=0$). But in practice, the inclusion of the transitional layer (resonant layer) modifies both the real part $\omega_r$ and the imaginary part $\gamma_{c0}$ of the wave frequency. In Fig.~\ref{dp_sssn1} (a) we show the $l/R$-dependent behavior of $\omega_r$ as a function of $k_zR$. As $l/R$ increases, $\omega_r$ shifts upward into a higher frequency, crossing over the frequency corresponding to $\omega_{Ci}(v_{Ci})$ at some value of $k_zR$. By crossing over it, it gets into the regime of the body modes and Eq.~(\ref{eq:24}) is no longer valid. We need to solve the connection formula for the body modes here. For the body mode, multiple eigenmodes and, as a result, multiple different damping rates for each $k_zR$ are obtainable. We plot one solution curve for each $l/R$ in the frequency regime of the body mode (above the line $v_{Ci}$) in the figure, by connecting the surface mode.

In Fig.~\ref{dp_sssn1} (b), we plot the damping rate $-\gamma_{c0}/\omega_r$ for $l/R=0.1,0.2,0.3,0.4$. For each value of $l$, the numerical solution has a local peak at a certain value of $k_zR$ while the analytical solution looks like a quadratic function of $k_zR$. A similar behavior of having a local maximum was found for the kink mode considering a linear density profile \citep{Soler2013}. As the value of $l/R$ becomes smaller, the peak position moves to higher values of $k_zR$ along with the increment of the maximum value of the damping rate. When $l/R=0.1$, the maximum value of the damping rate is $-\gamma_0/\omega_r\approx0.01$, which results in $\tau_D/T\approx14.11$. Although this ratio is large compared to the typical values (2-4) observed for the kink modes and the corresponding value of the analytical solution, it is not ignorable as previously expected and could be effective for wave damping. When the curve of phase speed crosses over the line $v_{Ci}$, the curve and the relevant damping rate correspond to body modes. As we have explained in Fig.~\ref{dp_sssn1} (a), in the body mode range $v_{Ci}<v<v_{si}$, multiple damping rates are obtainable. In the figure we plot only one solution curve of the body modes, which connects the surface sausage mode below the line $v_{Ci}$. We apply the same procedure to slow body kink modes.

In Fig.~\ref{dp_sssn1} (c), we plot the ratio of the damping time to the period $\tau_{D}/T(=1/(2\pi(-\gamma_{c0}/\omega_r)))$ for $l/R=0.1,0.2,0.3,0.4$. It has an inverse relation with the damping rate by its definition. It has a dip where the damping is most strong, which moves to the left as $l/R$ increases. As $l/R$ decreases, the minimum value (value at the dip) of $\tau_D/T$ gradually approaches about $10.5$.

\begin{figure}[ht]
\includegraphics[trim=0.0 85.0 25.0 5., clip, width=.5\textwidth]{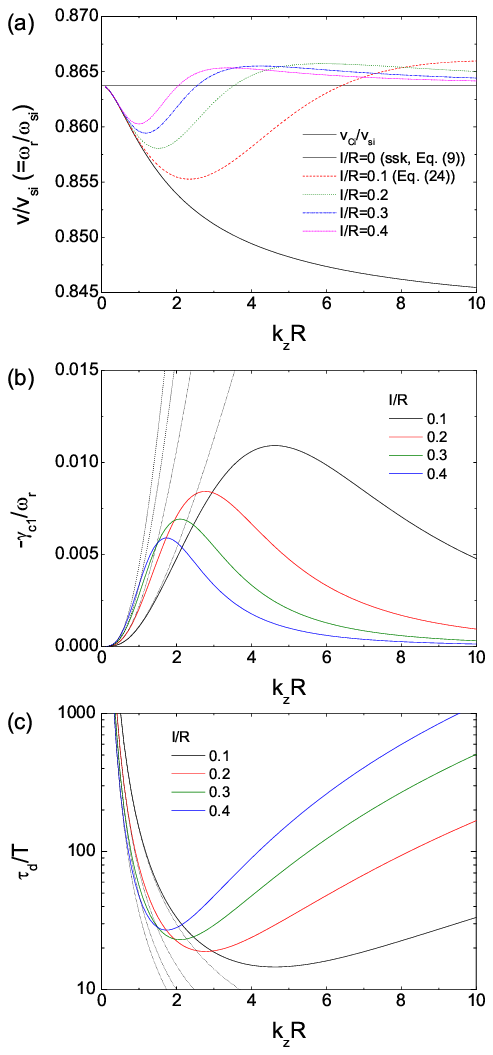}
\caption{\label{dp_ssksn1} (a) The phase speed of the slow surface kink (ssk) mode $v/v_{si}$ versus $k_zR$. We compare the solution of Eq.~(\ref{eq:9}) (solid line) without an inhomogeneous (transitional) layer with the solutions (dashed, dotted, dashed-dotted, and short-dashed lines) of Eq.~(\ref{eq:24}) with the transitional layer introduced in Sec.~\ref{sec4} when $l/R=0.1,0.2,0.3,0.4$. The other parameters are the same as in previous figures. The solution curve above the dotted line ($v_{Ci}$) correspond to the one of slow body kink modes. (b) The damping rate $-\gamma_{c1}/\omega_r$ versus $k_zR$. The analytical approximations (dotted lines), Eq.~(\ref{eq:33}), are compared to the numerical solutions (solid lines), Eq.~(\ref{eq:24}), for $m=1$. As $l/R$ increases the curve of the numerical solution shifts to the left while decreasing. The analytical and numerical solutions for each $l/R$ converge when the value of $k_zR$ approaches zero. (c) The ratio of damping time to period $\tau_D/T$ (logarithmic scale) versus $k_zR$. The features in (a), (b) and (c) are very similar to those of the slow surface sausage mode (Fig.~\ref{dp_sssn1}).}
\end{figure}
From the seismological point of view, the thickness of the transitional layer can be inferred from the damping time of the excited wave modes. So, the relation between damping rate (time) and the thickness is of interest. In Fig.~\ref{dp_sssn2}, we see the $l/R$ dependence of (a) the phase speed (eigenfrequency) $v/v_{si}$, (b) damping rate $-\gamma_{c0}/\omega_r$ and (c) the ratio of the damping time to the period $\tau_D/T$ by using Eq.~(\ref{eq:24}). When $k_zR$ is small the frequency monotonically increase, but for $k_zR>1$ it reaches a local peak and then decreases. For large $k_zR$ the wave frequency approaches $\omega_{Ci}(v_{Ci})$ as $l/R$ increases. The damping rate is in proportion to $l/R$ when $k_zR$ is small. As $k_zR$ increases the curve tends to have a local peak. The curve becomes sharper with an increment as $k_zR$ increases and the position of the peak shifts to smaller values of $k_zR$. This represents that the damping is efficient when $k_zR$ is large and $l/R$ is small. The ratio of damping time to the period reveals the opposite behavior to the damping rate as inferred from its definition. From the figure it is anticipated that $\tau_D/T$ could reach around 10 when the magnetic flux tube is very thin. From the behavior of the $l/R$-dependent damping rate, resonant absorption and the relevant damping of the slow surface sausage mode in the slow continuum would become significant for thinner transitional layers and for waves with small longitudinal wavelength.

We point out that for phase speeds larger than $v_{Ci}$ which corresponds to slow body sausage mode, one solution curve connected to the slow surface sausage mode is plotted for each $k_zR$ where one resonance point ($\delta=\delta_{c2}$) is considered as in the previous figure.  We postpone a detailed study on resonant absorption of slow body modes to the future.

\begin{figure}[ht]
\includegraphics[trim=5.0 85.0 10.0 5., clip, width=.5\textwidth]{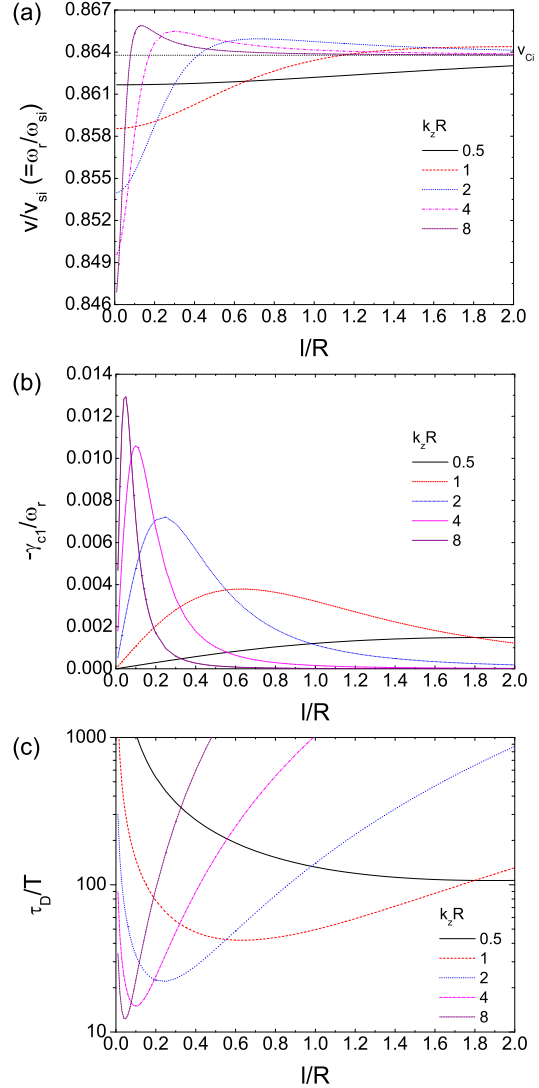}
\caption{\label{dp_ssksn2} (a) The phase speed of the slow surface kink (ssk) mode $v/v_{si}$ versus $l/R$. We show the $l/r$-dependent variation of the wave frequency when $k_zR=0.5,1,2,4,8$. The solution curve above the dotted line ($v_{Ci}$) correspond to the one of slow body kink modes. For larger $k_zR$, as $l/R$ increases the wave frequency approaches $\omega_{Ci}(v_{Ci})$. (b) The damping rate $-\gamma_{c1}/\omega_r$ versus $l/R$. Each curve has a local peak whose position shifts to a smaller $l/R$ as $k_zR$ increases, resulting in higher damping rate. (c) The ratio of damping time to the period $\tau_D/T$ (logarithmic scale) versus $l/R$. For sufficiently large value of $k_zR$, it has a local dip which moves to smaller $l/R$ as $k_zR$ increases. The features in (a), (b), and (c) are very similar to those of the slow surface sausage mode (Fig.~\ref{dp_sssn2}).}
\end{figure}

Together with the slow surface sausage (sss) mode, the slow surface kink (ssk) mode can undergo resonant absorption in the slow continuum. In Fig.~\ref{dp_ssksn1}, we plot the $k_zR$ dependence of the (a) phase speed, (b) damping rate, and (c) ratio of the damping time to the period for $l/R=0.1,0.2,0.3,0.4$. The deviation of the phase speed and damping rate from the $l/R=0$ case appears very similar to the case of the slow surface sausage mode. As $l/R$ increases the phase speed shifts upward crossing the line $v_{Ci}$ entering into the body mode range. The curve of the damping rate decreases as $l/R$ increases and the peak position moves to smaller $k_zR$ values. \citet{Soler2009} obtained a similar curve for the kink modes in solar filaments/prominence. They showed that for $l/R=0.2$, $\tau_D/T\approx1000$ as a minimum value, while our result gives it is about 19. The effect of the slow resonance on the wave damping is significant under photospheric conditions.

In Fig.~\ref{dp_ssksn2}, we plot the $l/R$ dependence of the (a) phase speed, (b) damping rate and (c) ratio of the damping time to the period for the slow surface kink mode when $k_zR=0.5,1,2,4,8$. All the features explained for Fig.~\ref{dp_sssn2} can apply here. The difference of resonant absorption in the slow continuum between the slow surface mode with $m=0$ and one with $m=1$ is small (see Fig.~\ref{dp_sk1}). Like as in previous two figures for the slow surface sausage modes, in Figs.~\ref{dp_ssksn1} and~\ref{dp_ssksn2}, the damping rate (time) for one of the slow body kink modes is plotted by connecting the slow surface kink mode when the phase speed is above $v_{Ci}$.

\begin{figure}[ht]
\includegraphics[trim=5.0 90.0 10.0 0., clip, width=.5\textwidth]{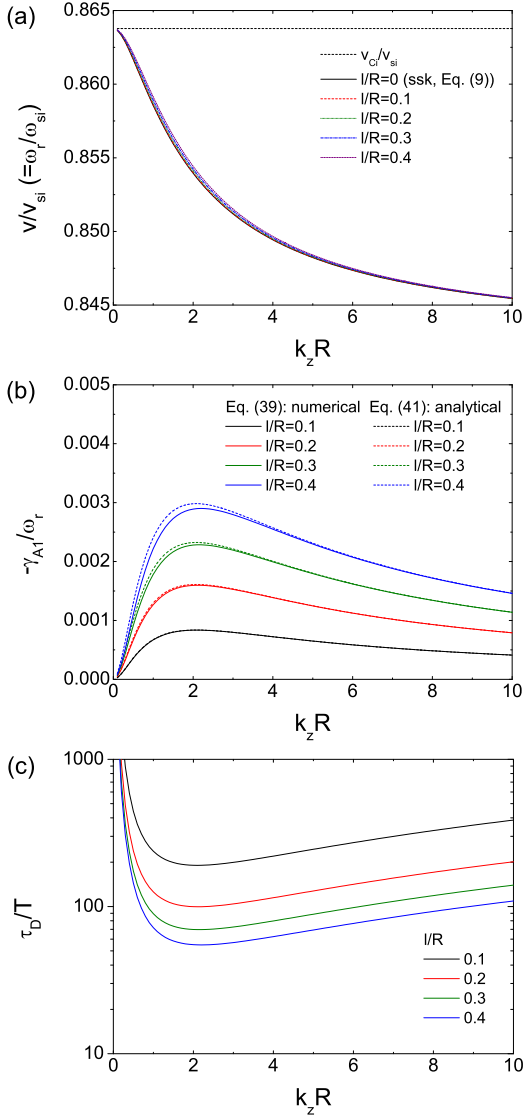}
\caption{\label{dp_sskn1} (a) The phase speed $v/v_{si}$ of the slow surface kink (ssk) mode versus $k_zR$. We compare the solution of Eq.~(\ref{eq:9}) (solid black line) without an inhomogeneous (transitional) layer with the solutions (dashed, dotted, dashed-dotted, short-dashed lines) of Eq.~(\ref{eq:39}) with the transitional layer introduced in Sec.~\ref{sec4} where $l/R=0.1,0.2,0.3,0.4$. The other parameters are the same as in previous figures. The wave frequency shifts upwards with little changes as $l/R$ increases, which is different from the sausage mode. (b) The damping rate $-\gamma_{A1}/\omega_r$ of the slow surface kink mode in the Alfv\'{e}n continuum versus $k_zR$. The analytical approximations (Eq.~(\ref{eq:41}), dashed lines) are compared to the numerical solutions (Eq.~(\ref{eq:39}), solid lines) for $m=1$. As $l/R$ increases the damping rate increases over all the range of $k_zR$ and the peak position moves to the right. The analytical solutions are well consistent with the numerical solutions where the deviation, which is still small, increases as $l/R$ increases. (c) The ratio of the damping time to the period $\tau_D/T$ (logarithmic scale) versus $l/R$: numerical calculations, Eq.~(\ref{eq:39}). }
\end{figure}

\begin{figure}[ht]
\includegraphics[trim=5.0 85.0 20.0 10., clip, width=.5\textwidth]{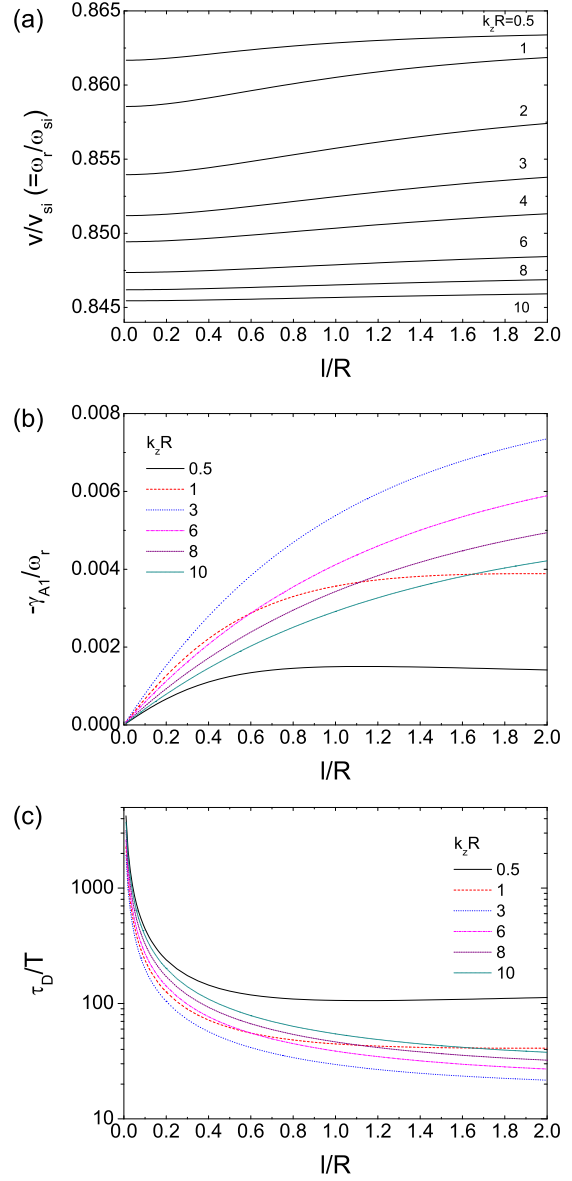}
\caption{\label{dp_sskn2} Numerical calculations of resonant absorption for the slow surface kink (ssk) mode in the Alfv\'{e}n continuum : Eq.~(\ref{eq:39}). (a) $v/v_{si}$ versus $l/R$ when $k_zR=0.5,1,2,3,4,6,8,10$. As $l/R$ increases the frequency shifts gradually upward. The $l/R$-dependent frequency shift is small, as for the $k_zR$ dependence. (b) $-\gamma_{A1}/\omega_r$ versus $l/R$ when $k_zR=0.5,1,3,6,8,10$  (numerical calculation, Eq.~\ref{eq:39}). Each curve increases monotonically as $l/R$ increases. For $k_zR<3$ the damping rate increases gradually in the whole range of $l/R$ and after $k_zR\approx3$ it decreases. (c) $\tau_D/T$ (logarithmic scale) versus $l/R$. The damping effect becomes significant as $l/R$ increases, while being most strong at $k_zR\approx3$.}
\end{figure}

While the slow surface sausage mode have no resonant absorption in the Alfv\'{e}n continuum since there is no azimuthal magnetic field in the equilibrium, we have two resonant absorptions for slow and fast surface kink modes in the Alfv\'{e}n continuum. In Fig.~\ref{dp_sskn1}, we show the (a) phase speed $v/v_{si}$, (b) damping rate $-\gamma_{c1}/\omega_r$, (c) ratio of the damping time to the period $\tau_D/T$ for the slow surface kink mode as a function of $k_zR$ when $l/R=0.1,0.2,0.3,0.4$. We use Eq.~(\ref{eq:39}) for numerical results and Eq.~(\ref{eq:41}) for analytical results. The wave frequency has little dependence on the $l/R$, slightly shifting upward as $l/R$ increases. When $l/R=0.1$ it is hard to distinguish from the original dispersion curve. The damping rate shows an increasing and then decreasing behavior having a local maximum (peak) at $k_zR\approx2$, similar to the behavior of the slow surface sausage mode. As $l/R$ increases, the damping rate increases in the whole range of $k_zR$ and the peak position shifts gradually to the right in the figure. It is worth to notice that the analytic results are very close to the numerical result, which means that the analytic formula, Eq.~(\ref{eq:41}), is a valid approximation for the resonant damping of the slow surface kink mode in the Alfv\'{e}n continuum in all the range of $k_zR$. The deviation of the numerical result from the analytical one becomes larger as $l/R$ increases. To compare with the slow surface sausage mode, the effect of the resonant damping on the slow surface kink mode looks much weaker than that of the slow surface sausage mode for small $l/R$ (See Fig.~\ref{dp_sk1} in more detail). The damping time over the period also appears to have a dip which goes down as $l/R$ increases, which is opposite to the two cases for the slow resonance. On the contrary, its behavior for the prominence reported by~\citet{Soler2009} is very different where for $l/R=0.2$ the damping time over the period was shown to not change from about 5 until $k_zR$ increases up to $0.1$, after which it increases rapidly as $k_zR$ increases. When $l/R=0.2$, we have $\tau_D/T\approx100$ as a minimum value. It was shown by~\citet{Soler2009} that in the solar filaments/prominences, the wave damping due to Alfv\'{e}n resonance is stronger than due to the slow resonance, which is reversed in the photospheric environment.

In Fig.~\ref{dp_sskn2}, we show the $l/R$-dependent behavior of the slow surface kink mode by numerical calculation (Eq.~(\ref{eq:39}) for $m=1$) when $k_zR=0.5,1,2,3,4,6,8,10$. It is found that (a) the phase speed has a small monotonic increment for each $k_zR$, similar to $k_zR$ dependence. In (b) the damping rate is shown to increase as a function of $l/R$. The damping rate first increases as $k_zR$ increases up to about 3 and then decreases again, in the whole range of $l/R$. It has a maximum value at $k_zR\approx3$. As a result, (c) the ratio of the damping time to the period $\tau_D/T$ has a minimum at $k_zR\approx3$. For the slow surface kink mode, it is expected that resonant absorption is most strong when $k_zR\approx3$ and $l/R$ is large. The value of $\tau_D/T$ reaches about $21.6$ when $k_zR=3$ and $l/R=2$.

{In our model configuration, the slow and Alfv\'{e}n continua do not overlap in the transitional layer. Hence when a wave is damped, it is either in the slow continuum or Alfv\'en continuum. There is no combination of the two resonant damping effects. If the two resonances overlap in the transitional layer, a combined effect could change the results for the slow surface kink mode, requiring further investigations.}

\begin{figure}[ht]
\includegraphics[trim=10.0 90.0 25.0 10., clip, width=.6\textwidth]{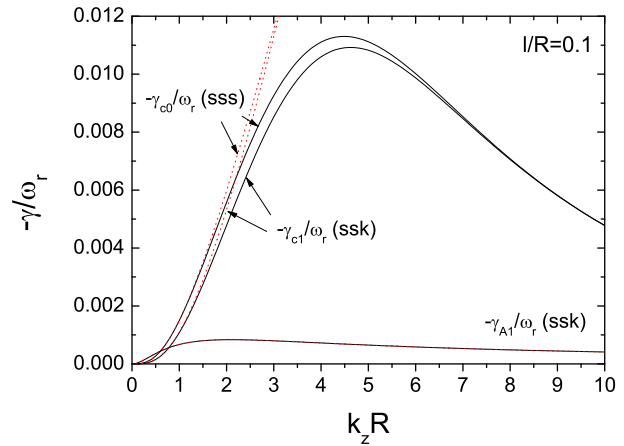}
\caption{\label{dp_sk1} Comparison of the damping rate $-\gamma_{c0}/\omega_r$, $-\gamma_{c1}/\omega_r$, and $-\gamma_{A1}/\omega_r$ as a function of $k_zR$ when $l/R=0.1$. Solid black lines represent numerical results while dotted red lines analytical results. For a small $k_zR$, both analytical approximation and numerical result show that the resonant damping of the slow surface kink (ssk) mode in the Alfv\'{e}n continuum is stronger than that of the slow surface sausage (sss) and kink (ssk) modes in the slow resonance. There is a crossover between damping rate due to the Alfv\'{e}n resonance and one due to slow resonance at certain values of $k_zR$. This feature is valid regardless of the value of $l/R$. The analytical calculations of slow surface sausage and kink modes undergoing the slow resonance converge to the numerical results when $k_zR$ goes to zero. On the contrary, the analytical solution of the slow surface kink mode undergoing the Alfv\'{e}n resonance is almost the same as the numerical result. }
\end{figure}

\begin{figure}[ht]
\includegraphics[trim=5.0 85.0 20.0 0., clip, width=.5\textwidth]{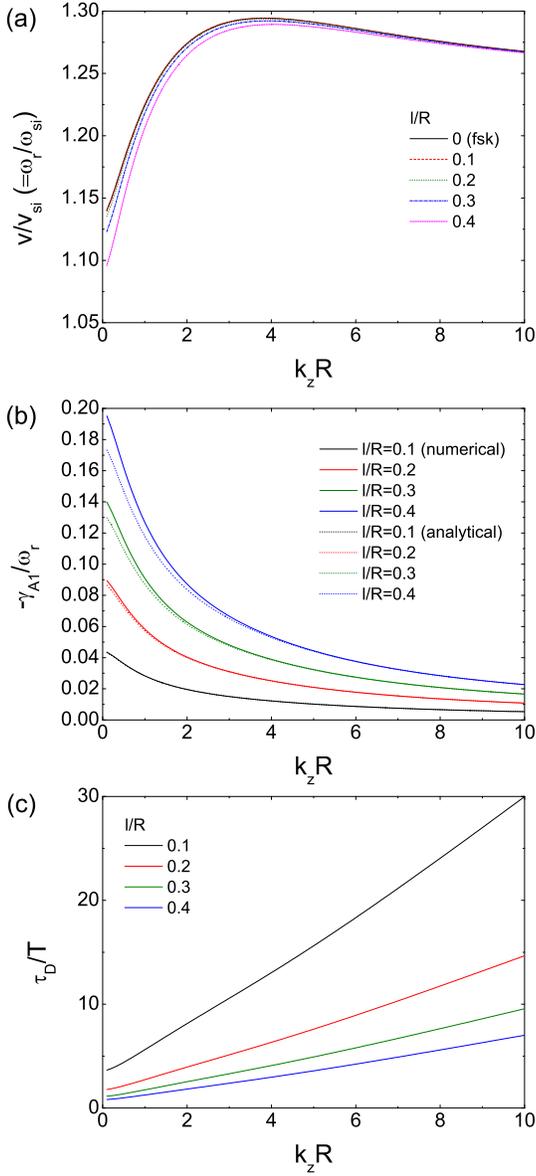}
\caption{\label{dp_fskn1} Numerical calculations of resonant absorption for the fast surface kink (fsk) mode in the Alfv\'{e}n continuum : Eq.~(\ref{eq:39}). (a) The phase speed $v/v_{si}$ of the fast surface kink (fsk) mode versus $k_zR$. We compare the solution of Eq.~(\ref{eq:9}) (fsk, solid black line) without an inhomogeneous (transitional) layer with the solutions (dashed, dotted, dashed-dotted, short-dahsed lines) of Eq.~(\ref{eq:39}) with the transitional layer introduced in Sec.~\ref{sec4} where $l/R=0.1,0.2,0.3,0.4$. The wave frequency shifts downwards with small changes as $l/R$ increases. The amount of shift is becomes larger as $k_zR$ approaches zero. These curves are convergent when $k_zR$ goes to infinity. (b) The damping rate $-\gamma_{A1}/\omega_r$ of the fast surface kink mode in the Alfv\'{e}n continuum versus $k_zR$. As $l/R$ increases the damping rate increases over the whole range of $k_zR$. Solid lines are obtained from Eq.~(\ref{eq:39}) and dotted lines from Eq.~(\ref{eq:41}). (c) The ratio of the damping time to the period $\tau_D/T$ versus $l/R$: numerical calculations, Eq.~(\ref{eq:39}).}
\end{figure}

We can think of the situation that two slow surface modes are excited simultaneously with the same amplitude since two modes are in the same frequency range.
In Fig.~\ref{dp_sk1}, we compare the three resonance effects: slow resonance on the slow surface sausage and slow surface kink modes and Aflv\'{e}n resonance on the slow surface kink mode when $l/R=0.1$. The two slow surface modes undergo a similar damping process in the slow resonance while the effect of resonant damping for slow surface kink mode in the Alfv\'{e}n continuum is quite small. As predicted from Eqs.~(\ref{eq:48}) and (\ref{eq:49}), when $k_zR\ll1$, the damping effect in the Alfv\'{e}n continuum is bigger than the damping effect in the slow continuum. As $k_zR$ increases the role of the two resonant effects is reversed and the difference increases until the damping rate in the slow continuum reaches a maximum. There is a crossover at $k_zR\approx0.6$ and $k_zR\approx0.78$. This feature maintains regardless of the value of $l/R$. If two slow modes are excited concurrently with a small longitudinal wavelength ($k_zR>1$), the slow surface kink mode would survive much longer than the slow surface sausage mode when ignoring other dissipation effects.  As inferred from Eq.~(\ref{eq:50}), it is also shown in the figure that the slow surface sausage mode is more easily damped than the slow surface kink mode in the slow continuum in the long wavelength limit.
\begin{figure}[ht]
\includegraphics[trim=5.0 90.0 25.0 10., clip, width=.5\textwidth]{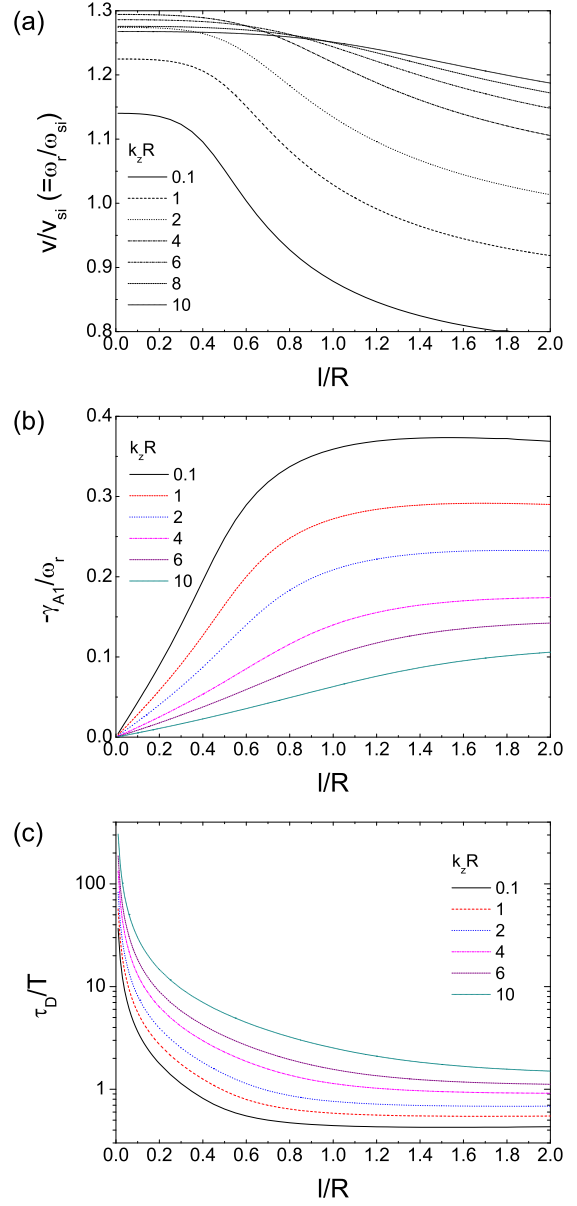}
\caption{\label{dp_fskn2} Numerical calculations for the fast surface kink (fsk) mode: Eq.~(\ref{eq:39}). (a) $v/v_{si}$ versus $l/R$. We show the $l/r$-dependent variation of the wave frequency when $k_zR=0.5,1,2,4,6,8,10$. The other parameters are the same as in previous figures. As $l/R$ increases the frequency shifts gradually downward. The $l/R$-dependent frequency shift is big for small $k_zR$ and decreases as $k_zR$ increases. (b) $-\gamma_{A1}/\omega_r$ versus $l/R$ when $k_zR=0.5,1,2,4,6,10$. Each curve increases monotonically as $l/R$ increases reaching a plateau for large $l/R$.  (c) $\tau_D/T$ (logarithmic scale) versus $l/R$. The damping effect becomes significant for smaller $k_zR$ and larger $l/R$.}
\end{figure}

Considering resonant absorption of the fast surface kink mode in the Alfv\^{e}n continuum, we find that the absorption behavior is different from that of the slow surface kink mode. It is shown in Fig.~\ref{dp_fskn1} (a) that the phase speed shifts downward as $l/R$ increases where $l/R=0.1,0.2,0.3,0.4$, which is opposite to the case of the slow surface kink mode. The value of the shift is small, but increases gradually as $l/R$ increases. In Fig.~\ref{dp_fskn1} (b), the damping rate approaches its maximum as $k_zR$ goes to zero and monotonically decreases as $k_zR$ becomes large from zero. This feature can be inferred from that the denominator of Eq.~(\ref{eq:41}) is proportional to $k_z^2R^2$ when $k_zR$ is small. The resonance effect for the fast surface kink mode is much bigger in comparison with the slow surface kink mode, leading to a strong wave damping like as under coronal conditions. The difference between analytical and numerical results grows proportionally to $l/R$, similar to the case of the slow surface kink mode.
In Fig.~\ref{dp_fskn1} (c), the ratio of the damping time to the period is shown to increase as $k_zR$ increases and as $l/R$ increases. The damping effect due to resonant absorption is most strong when $k_zR\approx0$ and $l/R$ is large.

In Fig.~\ref{dp_fskn2}, we present the $l/R$ dependence of the (a) phase speed, (b) damping rate, and (c) ratio of the damping time to the period when $k_zR=0.5,1,2,4,6,8,10$. The phase speed decreases as $l/R$ increases where the degree of change is big when $k_zR$ is small and becomes small as $k_zR$ increases. It also shows small deviations when $l/R$ is small, which means that the phase speed (or wave frequency) does not vary much for thin transitional layers. The behavior of the damping rate has a similar dependence on $l/R$ with the slow surface kink mode. There is no non-monotonic behavior with respect to $k_zR$, which appears for the slow surface kink mode. It is remarkable that the ratio of the damping time to the period can reach below 1 when $k_zR$ is small and $l/R$ is sufficiently large, which means that resonant absorption of the fast surface kink mode is very strong even under photospheric conditions although the result is based on the restricted assumption of thin transitional layers. The change of resonant absorption becomes small when $l/R>1$.


\section{Conclusion}
\label{sec6}

In a recent paper, we derived a general analytical formula (Eq.~(\ref{eq:28})) for the damping rate of the slow surface sausage mode in the slow continuum by considering the thin boundary (TB) approximation~\citep{Yu2017}. In this paper we have focused on resonant absorption both in the slow and Alfv\'{e}n continua under photospheric conditions, considering linear density and pressure (or squared magnetic field) profiles in the transitional layers. In order to study resonant absorption in the Alfv\'{e}n continuum we have applied the same procedure to obtain Eq.~(\ref{eq:28}) and derived another analytical formula, Eq.~(\ref{eq:40}).

In~\citet{Yu2017}, we have applied Eq.~(\ref{eq:28}) to the observational rapid damping of slow surface sausage mode in~\cite{Grant2015}. For the conventional magnetic pore $R\approx0.5-3Mm$ and $k_z=2\pi/\lambda_z=2\pi/4400km$, which yields $k_zR\approx0.7-4.3$. For $k_zR=4.3$ and $l/R=0.5$, our previous study based on the assumption of a linear cusp speed yields $-\gamma_{c0}/\omega_r=0.0089$ and $\tau_D/T\approx17.9$.
This value is reduced by a factor about 10 by using the linear density and linear pressure profiles in the transitional layer considered in this paper. This result implies that resonant absorption in the slow continuum could be efficient as a wave damping mechanism in the lower solar atmosphere. Another important point to mention is that resonant absorption is sensitive to the profiles of the physical quantities in the nonuniform layers. These analytical approximations predict that the damping rate increases as $l/R$ and $k_zR$ increase, but the numerical calculations show that it has a maximum value depending on both $l/R$ and $k_zR$. The peak position shifts toward smaller $k_zR$ values while decreasing its strength as $l/R$ increases. For example, when $l/R=0.1$, $-\gamma_{c0}/\omega_r=0.01128$, then $\tau_D/T\approx14.11$. Although this value seems quite big in comparison with the strong (rapid) damping, $\tau_D/T\simeq2-4$, of the fast kink modes in the Alfv\'{e}n continuum under coronal conditions, the resonant damping due to slow resonance could still be an efficient wave damping mechanism. Although we found that resonant absorption in the slow continuum is an efficient mechanism, this effect is too weak to explain the extremely rapid damping of the slow surface sausage mode observed by~\cite{Grant2015}. Other damping mechanism like, e.g., thermal conduction is needed.

The slow surface kink mode can resonantly damp both in the slow and Alfv\'{e}n continua. Its behavior in the slow continuum is very similar to the above features of the slow surface sausage mode. Therefore a similar wave damping due to resonant absorption in the slow continuum is expected for these two slow surface modes ($m=0,1$).

For resonant absorption in the Alfv\'{e}n continuum, it is found that the resonant damping manifests in a different way for each slow and fast surface kink modes. For the slow surface kink mode, the damping rate draws a curve as a function of $k_zR$ having a local maximum (peak) at a certain $k_zR$ and is proportional to $l/R$ regardless of the value of $k_zR$. The $l/R$ dependence of the damping time looks like following a power law when $l/R$ is small. The damping effect is most strong when $k_zR\approx3$ and $l/R=2$. When $k_zR=3$ and $l/R=2$, $-\gamma_{A1}/\omega_r\approx0.0735$, which gives $\tau_D/T\approx21.6$.
For the fast surface kink mode, the damping rate is a monotonically decreasing function of $k_zR$ and a monotonically increasing function of $l/R$. It becomes stronger as $k_zR$ goes to zero and $l/R$ increases. It is shown that the damping of the fast surface kink mode due to the resonance in the Alfv\'{e}n continuum could be very rapid in the photosphere as much as in the corona. For these kink modes, contrary to the slow surface sausage mode, the analytical approximations agree well with the numerical calculations.

Comparing resonant absorption of two slow surface modes, we could say that the strength of resonant absorption in the slow resonance is higher than that in the Alfv\'{e}n resonance except when $k_zR$ is very small. This relation is reversed as $k_zR$ increases. For a small value of $k_zR$ (long wavelength limit), we have derived analytical approximate formulas for three kink of resonant absorption and compared their relative strengths (Eqs.~(\ref{eq:48})-(\ref{eq:50})), which are well consistent with the numerical results.

Our study has dealt with only linear profiles for the density and pressure leaving a possibility of a higher damping rate for other certain profiles, for example, as shown by \citet{Soler2013} that linear, parabolic, and sinusoidal density profiles induce different behavior of damping rate for the kink mode under coronal conditions.

There is indeed a warning of using the obtained results for the thick transitional layers, as, e.g., ~\citet{VanDoorsselaere2004} pointed out that the thin tube thin boundary (TTTB) approximation induces significant deviation from exact numerical solutions up to 25\%  for the coronal loop oscillations. As we considered only an inhomogeneity in the radial direction, the stratification in the longitudinal direction~\citep[e.g.][]{Andries2005,Arregui2005,Dymova2006} or azimuthal direction may as well affect the resonant absorption behavior. Since the magnetic fluxes in the lower solar atmosphere are highly structured, the extension of the existing analytic approach of 1 dimensional resonant absorption to 2 or 3 dimensions is critical. Another subject we would mention is the resonant behavior of the body modes for $m=0,1$, which we leave as a future study.



\acknowledgments
 T.V.D. thanks the support from the Odysseus type II funding (FWO-Vlaanderen), IAP
P7/08 CHARM (Belspo), GOA-2015-014 (KU~Leuven), and European Research Council
(ERC) under the European Union's Horizon 2020 research and innovation
programme (grant agreement No 724326)

\appendix

\section{ Surface mode}
\label{append-a1}
For the surface mode with $m=1$ we have
\begin{eqnarray}
G_1&=&\frac{K_1(k_eR)}{K_1'(k_eR)}=\frac{-2K_1}{K_0+K_2},\label{eq:a1}\\
Q_1&=&\frac{I_1'(k_iR)K_1(k_eR)}{I_1(k_iR)K_1'(k_eR)}
=\bigg(\frac{I_0+I_2}{2I_1}\bigg)\bigg(\frac{-2K_1}{K_0+K_2}\bigg)\nonumber\\
&=&-\frac{K_1(I_0+I_2)}{I_1(K_0+K_2)},\label{eq:a2}\\
P_1&=&\bigg(\frac{I_1''}{I_1}-\frac{I_1'^2}{I_1^2}\bigg)\frac{K_1}{K_1'}\nonumber\\
&=&\bigg(\frac{3I_1+I_3}{4I_1}-\frac{(I_0+I_2)^2}{4I_1^2}\bigg)\frac{-2K_1}{K_0+K_2},\label{eq:a3}\\
S_1&=&\bigg(1-\frac{K_1''K_1}{K_1'^2}\bigg)\frac{I_1'}{I_1}\nonumber\\
&=&\bigg[1-\frac{(3K_1+K_3)K_1}{(K_0+K_2)^2}\bigg]\frac{I_0+I_2}{2I_1}.\label{eq:a4}
\end{eqnarray}

For the case $k_iR(k_eR)< 1$ (first order approximation) we derive
\begin{eqnarray}
G_1&\approx&\frac{\frac{1}{k_eR}-\frac{1}{4}+\frac{1}{2}[\ln(\frac{k_eR}{2})+\gamma_e]}
{-\frac{1}{(k_eR)^2}+\frac{1}{4}+\frac{1}{2}[\ln(\frac{k_eR}{2})+\gamma_e]}\nonumber\\
&\approx&-k_eR,\label{eq:a5}\\
Q_1&\approx&\bigg(\frac{1}{k_iR}+\frac{k_iR}{4}\bigg)
(-k_eR)=-\bigg(\frac{k_e}{k_i}+\frac{k_ik_eR^2}{4}\bigg),\label{eq:a6}\\
P_1&\approx&-k_eR\bigg(\frac{I_1''}{I_1}-\frac{I_1'^2}{I_1^2}\bigg)\nonumber\\
&\approx&-k_eR\bigg[\frac{\frac{3}{2}k_iR}{2k_iR}-\bigg(\frac{1}{k_iR}+\frac{k_iR}{4}\bigg)^2\bigg]\nonumber\\
&=&-k_eR\bigg[\frac{3}{4}-\bigg(\frac{1}{k_iR}+\frac{k_iR}{4}\bigg)^2\bigg]\nonumber\\
&\approx&-k_eR\bigg(\frac{1}{4}-\frac{1}{(k_iR)^2}\bigg),\label{eq:a7}\\
S_1&=&\bigg(1-\frac{K_1''K_1}{K_1'^2}\bigg)\frac{I_1'}{I_1}\approx\frac{1}{k_iR}\bigg(1-\frac{K_1''}{K_1'^2}\bigg)\nonumber\\
&=&\frac{1}{k_iR}\bigg\{1-\frac{\big(\frac{2}{(k_eR)^3}+\frac{1}{2k_eR}+\frac{k_eR[-5+12(\ln(k_eR/2)+\gamma_e)]}{32}\big)
}
{\big(-\frac{1}{(k_eR)^2}+\frac{1+2(\ln(k_eR/2)+\gamma_e)}{4}\big)^2}\nonumber\\
&&\times\bigg(\frac{1}{k_eR}+\frac{k_eR[-1+2(\ln(k_eR/2)+\gamma_e)]}{4}\bigg)\bigg\}\nonumber\\
&\approx&\frac{1}{k_iR}\bigg\{1-\frac{(2+\frac{(k_eR)^2}{2})(1+\frac{(k_eR)^2}{4}(-1+2f))}
{1-\frac{(k_eR)^2}{2}(1+2f)}\bigg\}\nonumber\\
&\approx&\frac{1}{k_iR}\bigg[1-\frac{2+f(k_eR)^2}
{1-\frac{(k_eR)^2}{2}(1+2f)}\bigg]\nonumber\\
&\approx&\frac{1}{k_iR}\bigg[1-(2+(k_eR)^2(1+3f))\bigg]\nonumber\\
&=&-\frac{1+[1+3(\ln(k_eR/2)+\gamma_e)](k_eR)^2}{k_iR}\nonumber\\
&\approx&-\frac{1+[1+3\ln(k_eR)](k_eR)^2}{k_iR},\label{eq:a8}
\end{eqnarray}
where $\gamma_e$ is the Euler's constant and, for $S_1$, $f=\ln(k_eR/2)+\gamma_e$ is used.

For the $m=0$ case, see Appendix A in~\citet{Yu2017}.

\section{Damping rate for the surface mode}
\label{append-2}
Here we briefly summarize the procedure to obtain the damping rate $-\gamma_m/\omega_r$~\citep[see][]{Yu2017}.
In order to calculate $\gamma_m$ we need to derive the expression for ${\partial D_{mr}}/{\partial \omega}$ where $\omega$ should be in the slow (cusp) or Alfv\'{e}n continuum.
We have
\begin{eqnarray}
\frac{\partial D_{mr}}{\partial \omega}&=&2\rho_i\omega-2\omega\rho_e\bigg(\frac{k_i}{k_e}\bigg)Q_m\nonumber\\
&&-\rho_e(\omega^2-\omega_{Ae}^2)\bigg(\frac{1}{k_e}\frac{d k_i}{d\omega}-\frac{k_i}{k_e^2}\frac{d k_e}{d\omega}\bigg)Q_m\nonumber\\
&&-\rho_e(\omega^2-\omega_{Ae}^2)\bigg(\frac{k_i}{k_e}\bigg)\frac{dQ_m}{d\omega}.\label{eq:b1}
\end{eqnarray}
For ${d k_i}/{d\omega}$ and ${d k_e}/{d\omega}$ we obtain
\begin{eqnarray}
\frac{d k_i}{d\omega}&=&-\frac{\omega^3}{v_{si}^2+v_{Ai}^2}\frac{(\omega^2-2\omega_{Ci}^2)}{(\omega^2-\omega_{Ci}^2)^2k_i},\label{eq:b2}\\
\frac{d k_e}{d\omega}&=&-\frac{\omega^3}{v_{se}^2+v_{Ae}^2}\frac{(\omega^2-2\omega_{Ce}^2)}{(\omega^2-\omega_{Ce}^2)^2k_e}.\label{eq:b3}
\end{eqnarray}
For ${dQ_m}/{d\omega}$ we obtain
\begin{eqnarray}
\frac{dQ_m}{d\omega}&=&R\bigg(\frac{I_m''}{I_m}-\frac{I_m'^2}{I_m^2}\bigg)\frac{K_0}{K_0'}\frac{dk_i}{d\omega}\nonumber\\
&&+R\bigg(1-\frac{K_m''K_m}{K_m'^2}\bigg)\frac{I_m'}{I_m}\frac{dk_e}{d\omega},\nonumber\\\label{eq:b4}
\end{eqnarray}
where the prime means the derivative with respect to the entire argument.

By means of Eqs. (\ref{eq:b2}) and (\ref{eq:b3}), Eq. (\ref{eq:b4}) becomes
\begin{eqnarray}
\frac{dQ_m}{d\omega}&=& \frac{k_iRP_m\omega^3(\omega^2-2\omega_{Ci}^2)}{(\omega^2-\omega_{si}^2)(\omega^2-\omega_{Ai}^2)(\omega^2-\omega_{Ci}^2)}
\nonumber\\&&+ \frac{k_eRS_m\omega^3(\omega^2-2\omega_{Ce}^2)}{(\omega^2-\omega_{se}^2)(\omega^2-\omega_{Ae}^2)(\omega^2-\omega_{Ce}^2)}\label{eq:b5}
\end{eqnarray}
where $P_m$ and $S_m$ are
\begin{eqnarray}
    P_m=\bigg(\frac{I_m''}{I_m}-\frac{I_m'^2}{I_m^2}\bigg)\frac{K_m}{K_m'},\label{eq:b6}\\
S_m=\bigg(1-\frac{K_m''K_m}{K_m'^2}\bigg)\frac{I_m'}{I_m}\label{eq:b7}.
\end{eqnarray}

Using Eqs.~(\ref{eq:b2}), (\ref{eq:b3}), and (\ref{eq:b5}) we have for $\partial D_{mr}/\partial\omega(=d D_{mr}/d\omega)$
\begin{eqnarray}
\frac{d D_{mr}}{d \omega}
&=&2\rho_i\omega-2\omega\rho_e\bigg(\frac{k_i}{k_e}\bigg)Q_m    
\nonumber\\
&&-\rho_e\omega^3(\omega^2-\omega_{Ae}^2)\bigg(\frac{k_i}{k_e}\bigg)\nonumber\\
&&\times
\frac{(\omega^2-2\omega_{Ci}^2)\big[Q_m+k_iR P_m\big]}{(\omega^2-\omega_{si}^2)(\omega^2-\omega_{Ai}^2)(\omega^2-\omega_{Ci}^2)}
\nonumber\\
&&+\rho_e\omega^3(\omega^2-\omega_{Ae}^2)\bigg(\frac{k_i}{k_e}\bigg)\nonumber\\
&&\times\frac{(\omega^2-2\omega_{Ce}^2)\big[Q_m-k_eR S_m\big]}{(\omega^2-\omega_{se}^2)
(\omega^2-\omega_{Ae}^2)(\omega^2-\omega_{Ce}^2)}.\nonumber\\\label{eq:b8}
\end{eqnarray}

Then the imaginary term $\gamma_m$ for the surface wave in the slow (cusp) continuum is

\begin{eqnarray}
\gamma_m&=&-{\frac{D_{mi}}{\frac{\partial D_{mr}}{\partial \omega}}\bigg|_{\omega=\omega_r}}\nonumber\\
&=&-\frac{\frac{\pi \rho_e k_z^2}{k_e\rho_c|\triangle_c|}\bigg(\frac{v_{sc}^2}{v_{sc}^2+v_{Ac}^2}\bigg)^2 
(\omega_r^2-\omega_{Ai}^2)(\omega_r^2-\omega_{Ae}^2)G_m}
{2\omega_r\big[1-\chi\big(\frac{k_i}{k_e}\big)Q_m\big]
-\omega_r\chi T_m},\nonumber\\\label{eq:b9}
\end{eqnarray}
where
\begin{eqnarray}
T_m&=&\omega_r^2(\omega_r^2-\omega_{Ae}^2)\bigg(\frac{k_i}{k_e}\bigg)\bigg\{
\frac{(\omega_r^2-2\omega_{Ci}^2)[Q_m+k_iR P_m]}{(\omega_r^2-\omega_{si}^2)(\omega_r^2-\omega_{Ai}^2)(\omega_r^2-\omega_{Ci}^2)}\nonumber\\
&&-\frac{(\omega_r^2-2\omega_{Ce}^2)[Q_m-k_eR S_m]}{(\omega_r^2-\omega_{se}^2)
(\omega_r^2-\omega_{Ae}^2)(\omega_r^2-\omega_{Ce}^2)}\bigg\}.\label{eq:b10}
\end{eqnarray}

Likewise, for the surface wave in the Alfv\'{e}n continuum, we obtain
\begin{eqnarray}
\gamma_m&=&-\frac{\frac{\pi \rho_e m^2}{k_e\rho_A|\triangle_A|r_A^2} 
(\omega_r^2-\omega_{Ai}^2)(\omega_r^2-\omega_{Ae}^2)G_m}
{2\omega_r\big[1-\chi\big(\frac{k_i}{k_e}\big)Q_m\big]
-\omega_r\chi T_m}.\label{eq:b11}
\end{eqnarray}

In the limit $k_iR(k_eR)\ll1$, $T_1$ becomes with the help of Eqs.~(\ref{eq:a5})-(\ref{eq:a8})
\begin{eqnarray}
T_1&=&-\omega_r^2(\omega_r^2-\omega_{Ae}^2)\bigg\{
\frac{(\omega_r^2-2\omega_{Ci}^2)(k_iR)^2}{2(\omega_r^2-\omega_{si}^2)(\omega_r^2-\omega_{Ai}^2)(\omega_r^2-\omega_{Ci}^2)}\nonumber\\&&
+\frac{(\omega_r^2-2\omega_{Ce}^2)[-(k_iR)^2/4+(1+3\ln(k_eR))(k_eR)^2]}{(\omega_r^2-\omega_{se}^2)
(\omega_r^2-\omega_{Ae}^2)(\omega_r^2-\omega_{Ce}^2)}\bigg\}.\nonumber\\\label{eq:b12}
\end{eqnarray}

For the slow surface kink mode with $k_zR\ll1$ ($\omega_r\approx\omega_{Ci}$), by the aid of Eqs.~(\ref{eq:12})-(\ref{eq:14}), Eq.~(\ref{eq:b12})  reduces to
\begin{eqnarray}
T_1&=&-\omega_{Ci}^2(\omega_{Ci}^2-\omega_{Ae}^2)\bigg\{
\frac{-\omega_{Ci}^2(k_iR)^2}{2(\omega_{Ci}^2-\omega_{si}^2)(\omega_{Ci}^2-\omega_{Ai}^2)\alpha}\nonumber\\&&
+\frac{(\omega_{Ci}^2-2\omega_{Ce}^2)[-(k_iR)^2/4+(1+3\ln(k_eR))(k_eR)^2]}{(\omega_{Ci}^2-\omega_{se}^2)
(\omega_{Ci}^2-\omega_{Ae}^2)(\omega_{Ci}^2-\omega_{Ce}^2)}\bigg\}\nonumber\\
&=&-\omega_{Ci}^2(\omega_{Ci}^2-\omega_{Ae}^2)\bigg\{
\frac{8\omega_{si}^2\omega_{Ai}^2}{\omega_{Ci}^8k_z^2R^2}
\bigg(\frac{\omega_{Ci}^2\omega_{Ai}^2}{\chi\omega_{si}^2(\omega_{Ci}^2-\omega_{Ae}^2)}-1\bigg)^2\nonumber\\&&
-\frac{1}{(\omega_{Ci}^2-\omega_{se}^2)(\omega_{Ci}^2-\omega_{Ae}^2)(\omega_{Ci}^2-\omega_{Ce}^2)}\nonumber\\&&
\times\bigg(\frac{\omega_{Ci}^2\omega_{Ai}^2}{\chi\omega_{si}^2(\omega_{Ci}^2-\omega_{Ae}^2)}-1\bigg)\nonumber\\&&
-\frac{(\omega_{Ci}^2-2\omega_{Ce}^2)(1+3\ln(k_zR))k_z^2R^2}{(\omega_{se}^2+\omega_{Ae}^2)(\omega_{Ci}^2-\omega_{Ce}^2)^2
}\bigg\}\nonumber\\
&\approx&-\frac{8\omega_{si}^2\omega_{Ai}^2(\omega_{Ci}^2-\omega_{Ae}^2)}{\omega_{Ci}^6k_z^2R^2}
\bigg(\frac{\omega_{Ci}^2\omega_{Ai}^2}{\chi\omega_{si}^2(\omega_{Ci}^2-\omega_{Ae}^2)}-1\bigg)^2.\nonumber\\&&
\label{eq:b13}
\end{eqnarray}

\section{Linear profiles for the density, squared magnetic field, and pressure}
\label{append-3}
For the linear profiles considered in Sec.~\ref{sec4} the variables $v_{s}$, $v_{A}$, and $v_{C}$ in the cusp resonance regime become
\begin{eqnarray}
v_{s}&=&\sqrt{\frac{\gamma p_c}{\rho_c}}=\sqrt{\frac{\gamma p_i}{\rho_i}}\sqrt{\frac{p_c}{p_i}}\frac{1}{\sqrt{\rho_{c}/\rho_i}}
\nonumber\\&=&\frac{v_{si}\sqrt{p_{ci}}}{\sqrt{1+\delta(\chi-1)}},\label{eq:d1}\\
v_{A}&=&\frac{B_c}{\sqrt{\mu_0\rho_c}}=\frac{B_i}{\sqrt{\mu_0\rho_i}}\frac{B_c}{B_i}\frac{1}{\sqrt{\rho_{c}/\rho_i}}\nonumber\\
&=&\frac{v_{Ai}B_{ci}}{\sqrt{1+\delta(\chi-1)}},\label{eq:d2}\\
v_{C}^2&=&\frac{v_{s}^2v_{A}^2}{v_{s}^2+v_{A}^2}=\frac{\frac{\gamma p_c}{\rho_c}\frac{B_c}{\sqrt{\mu_0\rho_c}}}{\frac{\gamma p_c}{\rho_c}+\frac{B_c^2}{\mu_0\rho_c}}\nonumber\\
&=&\frac{v_{si}^2v_{Ai}^2}{v_{si}^2p_{ci}+v_{Ai}^2B_{ci}^2}\frac{p_{ci}B_{ci}^2}{1+\delta(\chi-1)},\label{eq:d3}
\end{eqnarray}
where $\rho_{ci}=\rho_c/\rho_i$, $p_{ci}=p_c/p_i$ and $B_{ci}=B_{c}/B_{i}$. The subscript $c$ represents the value at the resonant position.
Assuming a linear variation of the squared magnetic field $B^2$ we can set $B_c^2=B_i^2+\delta(B_e^2-B_i^2)$ as for $\rho_c$, then
$p_c$ also has a similar relation $p_c=p_i+\delta(p_e-p_i)$ or vice versa. Making use of these variables Eqs.~(\ref{eq:d1})-(\ref{eq:d3}) reduce to

\begin{eqnarray}
v_{s}&=&\frac{v_{si}\sqrt{p_{ci}}}{\sqrt{1+\delta(\chi-1)}}=v_{si}\frac{\sqrt{1+\delta(p_{ei}-1)}}{\sqrt{1+\delta(\chi-1)}}\nonumber\\
&=&v_{si}\frac{\sqrt{1+\delta(\chi(v_{se}/v_{si})^2-1)}}{\sqrt{1+\delta(\chi-1)}}\nonumber\\
&=&v_{si}\frac{\sqrt{1+\delta(\chi v_{sei}^2-1)}}{\sqrt{1+\delta(\chi-1)}},\label{eq:d4}\\
v_{A}&=&=\frac{v_{Ai}B_{ci}}{\sqrt{1+\delta(\chi-1)}}=v_{Ai}\frac{\sqrt{1+\delta(B_{ei}^2-1)}}{\sqrt{1+\delta(\chi-1)}}\nonumber\\
&=&v_{Ai}\frac{\sqrt{1+\delta(\chi(v_{Ae}/v_{Ai})^2-1)}}{\sqrt{1+\delta(\chi-1)}}\nonumber\\
&=&v_{Ai}\frac{\sqrt{1+\delta(\chi v_{Aei}^2-1)}}{\sqrt{1+\delta(\chi-1)}},\label{eq:d5}\\
v_{C}^2&=&\frac{v_{s}^2v_{A}^2}{v_{s}^2+v_{A}^2}=\frac{v_{si}^2v_{Ai}^2}{v_{si}^2p_{ci}+v_{Ai}^2B_{ci}^2}
\frac{p_{ci}B_{ci}^2}{1+\delta(\chi-1)}\nonumber\\
&=&\frac{v_{si}^2v_{Ai}^2}{v_{si}^2[1+\delta(p_{ei}-1)]+v_{Ai}^2[1+\delta(B_{ei}^2-1)]}
\nonumber\\&&\times\frac{[1+\delta(p_{ei}-1)][1+\delta(B_{ei}^2-1)]}{1+\delta(\chi-1)},\label{eq:d6}\nonumber\\
&=&\frac{v_{si}^2v_{Ai}^2}{v_{si}^2[1+\delta(\chi v_{sei}^2-1)]+v_{Ai}^2[1+\delta(\chi v_{Aei}^2-1)]}
\nonumber\\&&\times\frac{[1+\delta(\chi v_{sei}^2-1)][1+\delta(\chi v_{Aei}^2-1)]}{1+\delta(\chi-1)},
\end{eqnarray}
where $v_{sei}^2=v_{se}^2/v_{si}^2$ and $v_{Aei}^2=v_{Ae}^2/v_{Ai}^2$.

From Eq. (\ref{eq:d3}) we derive the formula for $\delta(=\delta_c)$ with respect to $v_{C}$:
\begin{eqnarray}
A\delta^2+B\delta+C=0,\label{eq:d7}
\end{eqnarray}
where
\begin{eqnarray}
A&=&1+\frac{v_{C}^2}{v_{Ci}^2}(\chi-1)+\chi\bigg[\frac{v_{C}^2}{v_{Ci}^2}-(v_{sei}^2+v_{Aei}^2)\bigg]
\nonumber\\&&-\chi^2\bigg(\frac{v_{C}^2}{v_{Ci}^2}-v_{sei}^2v_{Aei}^2\bigg),\label{eq:d8}\\
B&=&2\bigg(\frac{v_{C}^2}{v_{Ci}^2}-1\bigg)-\chi\bigg[\frac{v_{C}^2}{v_{Ci}^2}
\bigg(1+\frac{v_{se}^2+v_{Ae}^2}{v_{si}^2+v_{Ai}^2}\bigg)\nonumber\\&&-(v_{sei}^2+v_{Aei}^2)\bigg],\label{eq:d9}\\
C&=&1-\frac{v_{C}^2}{v_{Ci}^2},\label{eq:d10}
\end{eqnarray}
which leads to two solutions:
\begin{eqnarray}
\delta_{c1}&=&-\frac{B}{2A}+\frac{\sqrt{B^2-4AC}}{2A}~~(0<\delta\leq\delta_m),\label{eq:d11}\\
\delta_{c2}&=&-\frac{B}{2A}-\frac{\sqrt{B^2-4AC}}{2A}~~(\delta_m\leq\delta\leq1),\label{eq:d12}
\end{eqnarray}
where $\delta_m$ is the value of $\delta$ when $v_C$ has a maximum value.

For $\triangle_c$ we obtain
\begin{eqnarray}
\triangle_c&=&d(\omega^2-\omega_{C}^2)/dr|_{r=r_c}=-2\omega_{Cc}\frac{d\omega_{Cc}}{dr}\nonumber\\
&=&-2\omega_{Cc}^2\bigg[\frac{v_{Ac}v_{sc}'+v_{sc}v_{Ac}'}{v_{sc}v_{Ac}}
-\frac{v_{sc}v_{sc}'+v_{Ac}v_{Ac}'}{v_{sc}^2+v_{Ac}^2}\bigg]\nonumber\\
&=&-\omega_{Cc}^2\bigg[\frac{p_c'}{p_c}-\frac{\rho_c'}{\rho_c}+2\frac{B_c'}{B_c}
-\frac{v_{sc}^2\big(\frac{P_c'}{P_c}\big)+2v_{Ac}^2\big(\frac{B_c'}{B_c}\big)}{v_{s}^2+v_{A}^2}\bigg]
\nonumber\\
&=&-\bigg(\frac{\omega_{Cc}^2}{l}\bigg)
\bigg\{\frac{(\chi v_{sei}^2-1)}{1+\delta(\chi v_{sei}^2-1)}-\frac{(\chi-1)}{1+\delta(\chi-1)}\nonumber\\&&+\frac{(\chi v_{Aei}^2-1)}{1+\delta(\chi v_{Aei}^2-1)}\label{eq:d13}\\&&-\frac{v_{si}^2(\chi v_{sei}^2-1)+v_{Ai}^2(\chi v_{Aei}^2-1)}{v_{si}^2[1+\delta(\chi v_{sei}^2-1)]+v_{Ai}^2[1+\delta(\chi v_{Aei}^2-1)]}\bigg\},\nonumber
\end{eqnarray}
\\
where the prime denotes the derivative with respect to $r$ and the subscript $c$ means $r=r_c$.

In the same way, we can derive $\delta(=\delta_a)$ in case of the Alfv\'{e}n resonance.
From Eq. (\ref{eq:d5}) we derive the formula for $\delta$ with respect to $v_{A}$
\begin{eqnarray}
\delta_a=\frac{1-(v_A/v_{Ai})^2}{1-(v_A/v_{Ai})^2+\chi[(v_A/v_{Ai})^2-v_{Aei}]}.~~~\label{eq:d14}
\end{eqnarray}

For $\triangle_A$ we obtain
\begin{eqnarray}
\triangle_A&=&d(\omega^2-\omega_{A}^2)/dr|_{r=r_A}=-2\omega_{Aa}\frac{d\omega_{Aa}}{dr}\nonumber\\ \nonumber\\
&=&-2\omega_{Aa}k_z\bigg(\frac{B_a}{\sqrt{\mu_0\rho_a}}\bigg)'
=-2\omega_{Aa}^2\bigg(\frac{B_a'}{B_a}
-\frac{1}{2}\frac{\rho_a'}{\rho_a}\bigg)\nonumber\\ \nonumber\\
&=&-\bigg(\frac{\omega_{Aa}^2}{l}\bigg)
\bigg\{\frac{\chi v_{Aei}^2-1}{1+\delta_a(\chi v_{Aei}^2-1)}-\frac{\chi-1}{1+\delta_a(\chi-1)}\bigg\},\nonumber\\\label{eq:d15}
\end{eqnarray}
where the subscript $a$ means $r=r_A~(\delta=\delta_a)$.

\bibliographystyle{aasjournal}
\bibliography{ramp_ref}


\end{document}